\documentclass[10pt,conference]{IEEEtran}
\usepackage{algorithm}
\usepackage{algpseudocode}

\usepackage[utf8]{inputenc} 
\usepackage[T1]{fontenc}
\usepackage{url}              
\usepackage{cite}             
\usepackage{algorithm}
\usepackage{algorithmicx}
\usepackage{algcompatible}
\usepackage{algpseudocode}

\usepackage[cmex10]{amsmath}  
\usepackage{mleftright}       
\mleftright                   

\usepackage{graphicx}         
\usepackage{booktabs}         

\usepackage{physics}
\usepackage{float}
\usepackage{caption}
\usepackage[utf8]{inputenc} 
\usepackage[T1]{fontenc}
\usepackage{url}
\usepackage{ifthen}
\usepackage{graphicx}
\usepackage{amsthm}
\usepackage{amsfonts}
\usepackage{cite}
\usepackage{multicol}
\usepackage{color}
\usepackage{amssymb}
\usepackage[cmex10]{amsmath} 
\usepackage{stfloats}
\usepackage{enumitem}
\newlist{steps}{enumerate}{1}
\setlist[steps, 1]{label = Step \arabic*:}
\usepackage{caption}

\newcommand{\un}[1]{\underline{#1}}

\newtheorem{defn}{Definition}
\newtheorem{thm}{{\cal T}heorem}
\newtheorem{cor}{Corollary}
\newtheorem{prop}{Proposition}
\newtheorem{lem}{Lemma}
\newtheorem{conj}{Conjecture}
\newtheorem{constr}{Construction}
\newtheorem{note}{Note}

\newcommand{\bit}{\begin{itemize}}
	\newcommand{\eit}{\end{itemize}}
\newcommand{\bcor}{\begin{cor}}
	\newcommand{\ecor}{\end{cor}}
\newcommand{\beq}{\begin{equation}}
	\newcommand{\eeq}{\end{equation}}
\newcommand{\beqn}{\begin{equation}}
	\newcommand{\eeqn}{\end{equation}}
\newcommand{\bea}{\begin{eqnarray}}
	\newcommand{\eea}{\end{eqnarray}}
\newcommand{\bean}{\begin{eqnarray*}}
	\newcommand{\eean}{\end{eqnarray*}}
\newcommand{\ben}{\begin{enumerate}}
	\newcommand{\een}{\end{enumerate}}
\newcommand{\bdefn}{\begin{defn}}
	\newcommand{\edefn}{\end{defn}}
\newcommand{\bnote}{\begin{note}}
	\newcommand{\enote}{\end{note}}
\newcommand{\bprop}{\begin{prop}}
	\newcommand{\eprop}{\end{prop}}
\newcommand{\blem}{\begin{lem}}
	\newcommand{\elem}{\end{lem}}
\newcommand{\bthm}{\begin{thm}}
	\newcommand{\ethm}{\end{thm}}
\newcommand{\bconj}{\begin{conj}}
	\newcommand{\econj}{\end{conj}}
\newcommand{\bconstr}{\begin{constr}}
	\newcommand{\econstr}{\end{constr}}
\newcommand{\bpf}{\begin{proof}}
	\newcommand{\epf}{\end{proof}}

\newcommand{\cc}{\mathcal{C}}

\newcommand{\baln}{\begin{align*}}
	\newcommand{\ealn}{\end{align*}}
\newcommand{\bal}{\begin{align}}
	\newcommand{\eal}{\end{align}}

\algdef{SE}[PARFOR]{ParFor}{EndParFor}[1]
{\algorithmicparallel\ \algorithmicfor\ #1\ \algorithmicdo}
{\algorithmicend\ \algorithmicparallel\ \algorithmicfor}
\algnewcommand{\algorithmicparallel}{\textbf{parallel}}

\usepackage{hyperref}
\usepackage[normalem]{ulem}

\hypersetup{
    colorlinks,
    citecolor=red,
    filecolor=black,
    linkcolor=blue,
    urlcolor=black
}

\usepackage{titlesec}

\usepackage{subcaption}
\usepackage{graphicx}
\usepackage{subcaption}

\IEEEoverridecommandlockouts
\begin{document}

\title{Impulse Decoding of Quantum LDPC Codes: Equivalence of Degeneracy and Code-Shortening}

\author{
\IEEEauthorblockN{Shobhit Bhatnagar, Michele Pacenti, Nithin Raveendran, David Declercq, and Bane Vasi\'c}
\IEEEauthorblockA{Department of Electrical \& Computer Engineering, The University of Arizona, Tucson, USA}

shobhitb97@gmail.com, mpacenti@arizona.edu, nithin@arizona.edu, declercq@codelucida.com, vasic@ece.arizona.edu
}
\maketitle
\begin{abstract} 
Quantum error correction is essential for building scalable quantum computers. Within the stabilizer formalism, the Calderbank–Shor–Steane framework constructs quantum codes from pairs of classical linear codes. A distinctive feature in this setting is \emph{degeneracy}, where multiple equivalent error estimates exist—a phenomenon that has no classical counterpart, and the lack of a meaningful classical coding-theoretic interpretation of which has remained a gap in the literature. In this paper, we demonstrate that degeneracy is closely related to the classical operation of shortening of a linear block code. Interestingly, the shortening here takes place at the decoder rather than at the encoder. Leveraging this insight, we present a parallel decoding scheme for quantum low-density parity-check codes, which we term \emph{impulse decoding}, that significantly outperforms belief propagation with ordered statistics decoding, as well as several other existing techniques, under both code-capacity and circuit-level noise, with significantly lesser complexity. We then present another algorithm based on decoding of residual errors, which when combined with impulse decoding achieves further performance improvement under circuit-level noise.
\end{abstract}

\section{Introduction}
\label{sec:shortening_Introduction}
Quantum computation promises exponential advantage over classical computation, however, quantum information is inherently more fragile than its classical counterpart. Quantum error correction (QEC) is therefore essential for building practical quantum computers. The most widely studied class of quantum error-correcting codes (QECCs) is that of stabilizer codes \cite{gottesman1997stabilizer}, which enable the construction of QECCs from classical error-correcting codes. In particular, the well-known family of Calderbank–Shor–Steane (CSS) codes \cite{calderbank1996good,steane1996error}, a subclass of stabilizer codes, allows one to construct a QECC from a pair of classical codes, where one contains the dual of the other.

Among the most promising candidates for achieving practical fault tolerance are quantum low-density parity-check (QLDPC) codes \cite{mackay2004sparse}, which are quantum analogues of classical LDPC codes \cite{gallager_ldpc}, and are typically constructed from pairs of classical LDPC codes within the CSS framework. Over the past decade, substantial effort has been devoted to designing QLDPC codes with both high rate and large minimum distance. This effort has resulted in several highly non-trivial code constructions \cite{hastings2021fiber,breuckmann2021balanced,panteleev2021quantum,tillich2013quantum}, and code families that simultaneously achieve asymptotically constant rate and fractional minimum distance are now known \cite{panteleev2022asymptotically,dinur2023good,leverrier2022quantum}.

One of the many reasons why QLDPC codes have gained a lot of traction in the literature is that, analogous to classical LDPC codes, they are equipped with efficient iterative decoding algorithms, such as belief propagation (BP) \cite{gallager_ldpc,aji2002generalized}, or its hardware-friendly variant, the min-sum algorithm (MSA) \cite{chen2005reduced}. One of the earliest works to study iterative decoding of QLDPC codes was carried out by Poulin et al. \cite{poulin2008iterative}. The authors in \cite{poulin2008iterative} highlighted two major differences between decoding in the classical and the quantum settings. First, unlike in the classical case where the Tanner graph of a code can be carefully designed, Tanner graphs associated with QLDPC codes must necessarily have short cycles (due to stabilizer commutativity constraints), which significantly impacts the performance of iterative decoders. Second, the phenomenon of degeneracy in QEC (discussed in detail later) gives rise to multiple equivalent error estimates, making it difficult for the decoder to converge to a single one. The phenomenon of degeneracy has no classical counterpart since a classical decoder needs to estimate the unique error vector produced by the channel. 
The lack of a clear classical coding-theoretic interpretation of degeneracy has remained a gap in the literature.

\emph{Our Contributions:}
In this paper, we demonstrate that degeneracy is closely related to the classical operation of shortening a linear block code.
Informally, shortening may be thought of as choosing a subcode of a linear code. We show that degeneracy allows the \emph{decoder} to shorten a linear block code. This is interesting since classically, the operation of shortening can only be performed at the encoder, not at the decoder. 

\begin{figure}[ht]
    \centering
    \includegraphics[width=0.9\linewidth]{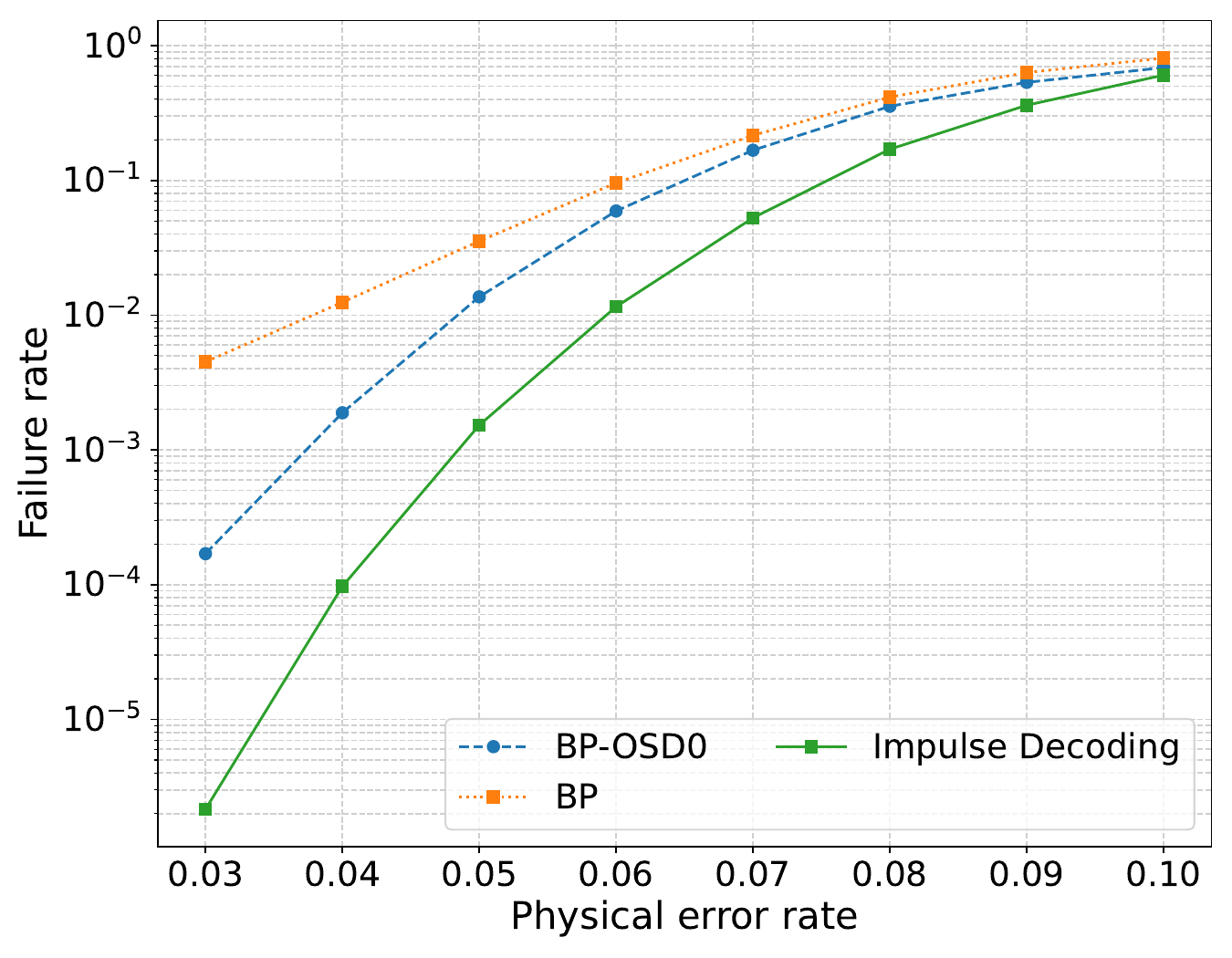}
    \caption{Performance of impulse decoding (Algorithm \ref{algo:impulse_decoder}) for the $[[288,12,18]]$ BB code under code-capacity noise.
    }
    \label{fig:BB288_1}
\end{figure}

Next, by leveraging the connection to code-shortening, we present a parallel decoding scheme for QLDPC codes. We refer to this scheme as \emph{impulse decoding}, since shortening can be realized in the BP framework by setting the value of log-likelihood ratios of variable nodes to infinite. Simulation results for the popular code families of bivariate-bicycle (BB) codes \cite{bravyi2024high} and lifted product  (LP) codes \cite{panteleev2021degenerate} demonstrate that impulse decoding achieves state-of-the-art performance, significantly outperforming benchmark decoders such as BP with ordered statistics decoding (BP-OSD) \cite{panteleev2021degenerate,roffe2020decoding} under both code-capacity and circuit-level noise. Moreover, depending on the degree of parallelization, the impulse decoder requires minimal post-processing, leading to a substantial reduction in complexity. The performance of impulse decoding (Algorithm \ref{algo:impulse_decoder} below) for the $[[288,12,18]]$ BB code under code-capacity noise is presented in Fig. \ref{fig:BB288_1}. We also present and discuss several interesting observations and insights from our simulations that highlight the close interplay between degeneracy and decoding dynamics, and their impact on both decoding performance and latency. Decoding QLDPC codes has been an active area of research since the work of Poulin et al. \cite{ducrest2024check,du2022stabilizer,hillmann2024localized,demarti2024almost,demarti2024closed,wolanski2024ambiguity,delfosse2021almost,bhattacharyya2025decoding,yin2024symbreak,hack2024belief,ninkovic2024decoding,raveendran2021trapping}, and
many recently-proposed techniques \cite{valentini2025restartbeliefgeneralquantum,ye2025beamsearchdecoderquantum,muller2025improved,gong2024toward,yao2024belief,alinia2025decimation,tsubouchi2025degeneracy} also proceed by fixing variable nodes, though largely based on heuristic arguments or empirical observations. The connection to code-shortening established here provides a more principled basis for such an approach, and also highlights how degeneracy can be exploited more directly to improve decoding performance and latency. 

We then present another decoding algorithm that combines impulse decoding with a residual-error-based decoding approach. This achieves further performance improvement under circuit-level noise while requiring a smaller number of parallel decoders, at the expense of performing serial processing within each decoder.

\textit{Organization of the paper:} We begin in Section \ref{sec:shortening_preliminaries} by reviewing the stabilizer formalism for QEC, the shortening operation, the BP algorithm and circuit-level noise. In Section \ref{sec:degeneracy_and_shortening}, we establish the connection between degeneracy and code-shortening. Section \ref{sec:Impulse_Decoder} introduces the impulse decoder and its variants, and Section \ref{sec:Numerical_Results} presents simulation results. Finally, Section \ref{sec:conclusion} concludes the paper.

\section{Preliminaries}
\label{sec:shortening_preliminaries}
\emph{Notation:} For integers $m,n$, where $m<n$, we use $[m:n]$ to denote the set $\{m,m+1,\dots,n\}$  
The set $[1:n]$ will sometimes also be denoted by $[n]$. 
We use $I_n$ to denote the $(n\times n)$ identity matrix. The zero vector of length $n$ is denoted by $\un{0}_n$.
When the length is clear from context, we omit the subscript and write $\un{0}$ to denote a zero vector of appropriate length. Similarly, the $(m \times n)$ zero matrix is denoted by $\un{\un{0}}_{m \times n}$, and, when the dimensions are clear from context, simply by $\un{\un{0}}$. For a vector $\un{e}\in \mathbb{F}_2^n$, where $\mathbb{F}_2$ denotes the finite field of two elements, we use ${e}_i$ to denote the $i$-th coordinate of $\un{e}$ 
and $\un{e}_\mathcal{I}$ to denote the sub-vector of $\un{e}$ indexed by the coordinates in $\mathcal{I}\subseteq [n]$. We use $\text{supp}(\un{e})$ to denote the support of $\un{e}$, i.e., $\text{supp}(\un{e})=\{i \in [n]\mid {e}_i\ne 0\}$. The notation $|\mathcal{I}|$ denotes that cardinality of a set $\mathcal{I}$, i.e., the number of elements in $\mathcal{I}$.
The Hamming weight of a vector $\un{e}$ is denoted by $\text{wt}(\un{e})$ and is the size of its support.
For a matrix $A \in \mathbb{F}_2^{m \times n}$, we denote its $i$-th column by $A_i$ and its $j$-th row by $A_{\text{row},j}$. 
We will use $P\otimes Q$ to denote the tensor product of two matrices $P \in \mathbb{C}^{m_1\times n_1}$ and $Q \in \mathbb{C}^{m_2\times n_2}$, where $\mathbb{C}$ denotes the scalar field of complex numbers. 

\subsection{Shortening of a Linear Code}
Let $\mathcal{C}$ be an \([n,k,d]\) linear block code, and let $\mathcal{I}\subseteq[n]$ be a subset of indices.
The code $\mathcal{C}$ shortened on the coordinates in $\mathcal{I}$, denoted by $\mathcal{C}_{\mathcal{I}}$, refers to the subcode (subspace) of $\cc$ obtained by collecting the codewords of $\mathcal{C}$ that are zero on all coordinates in \( \mathcal{I} \) \cite{huffman2010fundamentals}. More formally,
$$
\mathcal{C}_{\mathcal{I}} =
\{\ \un{c} \in \mathcal{C} \mid \un{c}_i = 0 \ \forall i \in \mathcal{I}\}.
$$
In the standard definition of shortening \cite{huffman2010fundamentals}, one deletes the coordinates $i\in \mathcal{I}$ from all the codewords of $\mathcal{C}_{\mathcal{I}}$ (since they are known to be $0$), however, for convenience of notation, here we will retain these coordinates.
Shortening can therefore be interpreted as selecting the subset of codewords of $\mathcal{C}$ whose entries indexed by $\mathcal{I}$ are all $0$. In this paper, we will use the term shortening in the more general sense of choosing a subset of codewords whose entries on an index set $\mathcal{I}$ are fixed to some element $\un{u}$ of $\mathbb{F}_2^{|\mathcal{I}|}$. In this case, we will refer to the code as being \emph{shortened on $\mathcal{I}$ to $\un{u}$}. In other words, the code $\mathcal{C}$ shortened on $\mathcal{I}$ to $\un{u}$ is the set 
\begin{equation*}
\label{eq:general_shortening_def}
    C_{\mathcal{I},\un{u}} =
\{\, \un{c} \in \mathcal{C} \mid \un{c}_\mathcal{I} = \un{u} \}.
\end{equation*}
For the case when $\mathcal{I}=\{i\}$ is a singleton set, we will say that the code $\cc$ is shortened on $i$ to $0$, if $u=0$, or to $1$, if $u=1$. Note that, by definition, ${C}_{\mathcal{I},\un{0}} = {C}_{\mathcal{I}}$.
Also note that $C_{\mathcal{I},\un{u}}$ may not be a subcode (subspace) of $\mathcal{C}$ in general. 
Consider a codeword $\un{c}\in\mathcal{C}$ and an index set $\mathcal{I}\subseteq[n]$, and let $\un{u}=\un{c}_\mathcal{I}$. Since ${C}_{\mathcal{I},\un{0}}$ is a subcode of $\mathcal{C}$, we have the coset $\{\un{c}+\un{c}'\mid \un{c}'\in\mathcal{C}_{\mathcal{I},\un{0}}\}$ of $\mathcal{C}_{\mathcal{I},\un{0}}$ in $\mathcal{C}$. It is easy to see that $C_{\mathcal{I},\un{u}}$ is precisely this coset. Thus, for any $\mathcal{I}\subseteq[n]$ and any $\un{u} \in \mathbb{F}_2^{|\mathcal{I}|}$, we have that $C_{\mathcal{I},\un{u}}$ is either empty, or it is a coset of $\mathcal{C}_{\mathcal{I},\un{0}}$ in $\mathcal{C}$. It is easy to see that if the columns of a generator matrix of $\mathcal{C}$ indexed by $\mathcal{I}$ are linearly independent, then ${C}_{\mathcal{I},\un{u}}$ is non-empty for every $\un{u}\in\mathbb{F}_2^{|\mathcal{I}|}$. 

\subsection{Stabilizer Formalism of QEC}
As mentioned before, among the most powerful and widely used constructions of QECCs are stabilizer codes \cite{gottesman1997stabilizer}, which we now briefly describe. We begin by introducing the single-qubit Pauli operators. These are the complex matrices $I_2,X,Y$ and $Z$, where
\begin{equation*}
X =
\begin{pmatrix}
0 & 1 \\
1 & 0
\end{pmatrix}, \quad
Y =
\begin{pmatrix}
0 & -\omega \\
\omega & 0
\end{pmatrix}, \quad
Z =
\begin{pmatrix}
1 & 0 \\
0 & -1
\end{pmatrix}.
\end{equation*}
Here, $\omega = \sqrt{-1}$ denotes the imaginary unit.
Errors acting on an $n$-qubit system are described by tensor products of single-qubit Pauli operators. The Pauli group on $n$ qubits, denoted $\mathcal{P}_n$, is given by
\begin{equation*}
\mathcal{P}_n =
\{
\omega^k P_1 \otimes \cdots \otimes P_n
\mid
P_i \in \{I_2, X, Y, Z\},\ 
k \in [0:3]
\}.
\end{equation*}
The group $\mathcal{P}_n$ is non-abelian but has the important property that any two elements either commute or anti-commute (which follows from the properties of single-qubit Pauli operators described above). Global phases $\{\pm 1, \pm \omega\}$ play no physical role and are typically modded out when discussing error correction. The weight of an element of $\mathcal{P}_n$ is defined to be the number of non-identity components in the tensor product.

The stabilizer formalism describes a quantum code as the common eigenspace of a set of commuting $n$-qubit Pauli operators. A \emph{stabilizer group} $\mathcal{S}$ is an Abelian subgroup of $\mathcal{P}_n$ such that $-I_2^{\otimes n} \notin \mathcal{S}$. Let $\mathcal{S}$ be generated by $S_1, S_2, \dots, S_{n-k},$ i.e.,
$$
\mathcal{S} = \langle S_1, S_2, \dots, S_{n-k} \rangle,
$$
where the generators $S_i \in \mathcal{P}_n$ commute pairwise and are independent.
The associated stabilizer code $\mathcal{Q}_\mathcal{S}$ is defined as the simultaneous $+1$-eigenspace of all stabilizer elements,
\begin{equation*}
\mathcal{Q}_\mathcal{S}
=
\{
\ket{\psi} \in \mathbb{C}^{2^n}
\mid 
S \ket{\psi} = \ket{\psi} \;\; \forall S \in \mathcal{S}
\},
\end{equation*}
where we have used Dirac's bra-ket notation to denote state vectors.
If $\mathcal{S}$ has $n-k$ independent generators, then $\mathcal{Q}_\mathcal{S}$ has dimension $2^k$ and encodes $k$ logical qubits into $n$ physical qubits.

The \emph{centralizer} of $\mathcal{S}$ in $\mathcal{P}_n$ is given by
$\mathcal{C}(\mathcal{S})
=
\left\{
E \in \mathcal{P}_n
\;\middle|\;
ES = SE \;\; \forall S \in \mathcal{S}
\right\},$
i.e., $\mathcal{C}(\mathcal{S})$ consists of the elements of $\mathcal{P}_n$ that commute with every element of $\mathcal{S}$. Since $\mathcal{S}$ is abelian by construction, $\mathcal{S} \subseteq \mathcal{C}(\mathcal{S})$. 
When a Pauli error $E \in \mathcal{P}_n$ acts on a code state $\ket{\psi}\in \mathcal{Q}_\mathcal{S}$, the syndrome, which is obtained by measuring the stabilizer generators, depends entirely on the commutation relations between $E$ and the stabilizer generators.
If $E$ commutes with all stabilizer generators, i.e., if $E \in \mathcal{C}(\mathcal{S})$, the syndrome is trivial, making the error undetectable. Since the elements of the stabilizer group $\mathcal{S}$ act as the identity on the code space, the set of harmful errors is given by $\mathcal{C}(\mathcal{S})\setminus\mathcal{S}$.
The minimum distance of the stabilizer code is therefore defined as the minimum weight of an element of $\mathcal{C}(\mathcal{S})\setminus\mathcal{S}$.

Two Pauli errors \(E\) and \(E'\) yield the same syndrome if and only if \(E' = EC\) for some \(C \in \mathcal{C}(\mathcal{S})\). The syndrome thus identifies only the coset \(E\mathcal{C}(\mathcal{S})\). Since $\mathcal{C}(\mathcal{S})$ may be further decomposed as \(\mathcal{C}(\mathcal{S}) = \bigcup_{\ell} L_{\ell}\mathcal{S}\), this leaves ambiguity among cosets \(E L_{\ell}\mathcal{S}\). Errors differing by stabilizers act identically, while those differing by logical operators do not. Thus, decoding amounts to selecting the correct \(\mathcal{S}\)-coset within \(E\mathcal{C}(\mathcal{S})\). The fact that any \(ES\), \(S \in \mathcal{S}\), suffices to correct \(E\) is called \emph{degeneracy}. Degeneracy is unique to quantum error correction, since in a classical setting, the decoder must identify exactly the error that is introduced by the channel, resulting in a unique correct error estimate. 

\subsection{Calderbank--Shor--Steane (CSS) Codes}

CSS codes \cite{calderbank1996good,steane1996error} form an important subclass of
stabilizer codes that are constructed from a pair of classical binary
linear codes. Let
$
\mathcal{C}_X, \mathcal{C}_Z \subseteq \mathbb{F}_2^n
$
be two classical linear codes satisfying the condition
$
\mathcal{C}_X^\perp \subseteq \mathcal{C}_Z ,
$
where \(\mathcal{C}_X^\perp\) denotes the dual of \(\mathcal{C}_X\) under
the standard binary inner product.
Let \(H_X \in \mathbb{F}_2^{r_X \times n}\) and \(H_Z \in \mathbb{F}_2^{r_Z \times n}\) be parity-check matrices for \(\mathcal{C}_X\)
and \(\mathcal{C}_Z\), respectively.
The stabilizer group of the corresponding CSS code $\text{CSS}({\mathcal{C}_X,\mathcal{C}_Z})$ is generated by
purely \(X\)-type and purely \(Z\)-type operators (corresponding to the rows of $H_X$ and $H_Z$, respectively), which may be
represented compactly by the block stabilizer matrix
\begin{equation*}
H_{\text{CSS}} =
\begin{pmatrix}
H_X & 0 \\
0   & H_Z
\end{pmatrix}.
\end{equation*}
The relation \(\mathcal{C}_X^\perp \subseteq \mathcal{C}_Z\) is
equivalent to the matrix orthogonality condition
$
\label{eq:css_condition}
H_X H_Z^T = 0,
$
 which is precisely the symplectic commutativity condition, thus
ensuring that all stabilizers commute. In the special case when both $H_X$  and $H_Z$ are sparse matrices, the corresponding CSS code is called a QLDPC code, in analogy with classical LDPC codes.

The number of logical qubits is given by
$k = n - \mathrm{rank}(H_X) - \mathrm{rank}(H_Z).$
Logical $Z$ operators correspond to  $\mathcal{C}_X / \mathcal{C}_Z^\perp$, while logical $X$ operators correspond to $\mathcal{C}_Z /\mathcal{C}_X^\perp$. In other words $\mathcal{C}_X$ and $\mathcal{C}_Z$ have generator matrices $G_X$ and $G_Z$, respectively, of the form
\begin{equation*}
G_X =
\begin{pmatrix}
H_Z \\
L_Z
\end{pmatrix},
\qquad
G_Z =
\begin{pmatrix}
H_X \\
L_X
\end{pmatrix}.
\end{equation*}
The rows of \(L_X\) and \(L_Z\) represent nontrivial logical \(X\) and
logical \(Z\) operators, respectively. The distance of $\text{CSS}({\mathcal{C}_X,\mathcal{C}_Z})$ is given by  
\begin{equation}
\label{eq:dmin_of_CSS}
d = \min \left\{
\min_{\,\un{z} \in \mathcal{C}_X \setminus \mathcal{C}_Z^\perp} \mathrm{wt}(\un{z}),
\;
\min_{\,\un{x} \in \mathcal{C}_Z \setminus \mathcal{C}_X^\perp} \mathrm{wt}(\un{x})
\right\}.
\end{equation}

For an error $\un{e} = (\un{e}_X \mid \un{e}_Z)^T$, the syndrome $\un{s}$ is obtained by taking the symplectic inner product of $\un{e}$ with each row of $H$. Thus, $\un{s}$ has the form $\un{s}=(\un{s}_Z, \un{s}_X)^T,$ where $\un{s}_X = H_Z \un{e}_X$ and $\un{s}_Z = H_X \un{e}_Z$.
When $X$-type and $Z$-type errors occur independently, they can be corrected separately using the classical codes defined by $H_X$ and $H_Z$. In the rest of this paper, we will only focus on correcting $X$-type errors, $Z$-type errors can be handled analogously. We will therefore drop the subscripts and use the notation $\un{e}\in\mathbb{F}_2^n$ for an $X$-type error, and $\un{s}=H_Z\un{e}$ for the corresponding syndrome.
Without loss of generality, we will assume that $H_X$ is of the form $H_X = [I_\rho \mid A]$, and $G_Z$ is of the form 
\begin{equation}
\label{eq:Gz_form}
G_Z =
\begin{pmatrix}
I_{\rho} & A \\
0   & L_X
\end{pmatrix}.
\end{equation}

\subsection{Syndrome Decoding of QLDPC Codes}
Belief propagation is a well-known low-complexity iterative decoding algorithm for classical LDPC codes \cite{gallager_ldpc,richardson2008modern}. It is a message-passing algorithm that passes messages along the edges of the Tanner graph of an LDPC code. Let $H \in \mathbb{F}_2^{m \times n}$ be a parity-check matrix of a binary $[n,k \geq (n-m)]$ LDPC code. The corresponding Tanner graph $\mathcal{T}(H) = (\mathcal{V}, \mathcal{C}, \mathcal{E})$ is a bipartite graph, where $\mathcal{V} = \{v_1,v_2,\dots, v_n\}$ is the set of variable nodes corresponding to the $n$ code symbols and $\mathcal{C} = \{c_1,c_2, \dots, c_m \}$ is the set of check nodes that correspond to the rows of $H$. 
There is an edge connecting check node $i$ and variable node $j$ iff $H_{i,j} = 1$, where $H_{i,j}$ denotes the $(i,j)$-th entry of $H$. 
The neighborhood $\mathcal{N}(v)$ of a variable node $v \in \mathcal{V}$ is the set of all check nodes that are connected to it. Similarly, the neighborhood $\mathcal{N}(c)$ of a check node $c \in \mathcal{C}$ is the set of all variable nodes that are connected to it. 

We now come to the quantum setting. Consider a CSS code $\text{CSS}({\mathcal{C}_X,\mathcal{C}_Z})$, and a quantum channel that generates an $X$-type error on each qubit with probability $p$, independent of other qubits. We obtain the syndrome $\un{s} = H_Z\un{e}$ as mentioned in the previous section.
The goal of BP decoding is to find an error estimate $\hat{\un{e}}$ such that $H_Z\hat{\un{e}} =\un{s}$. 

Message-passing is performed on the Tanner graph $\mathcal{T}(H_Z) = (\mathcal{V}, \mathcal{C}, \mathcal{E})$ corresponding to $H_Z$. 
The messages exchanged between check and variable nodes are in the form of log-likelihood ratios (LLRs). Since $X$-type errors occur with probability $p$, the channel LLR associated with each variable node $v \in \mathcal{V}$ is given by
$$\lambda_{v} = \ln\left(\frac{1-p}{p}\right).$$

At the $0^\text{th}$ iteration $t = 0$, each variable node $v \in \mathcal{V}$ passes the message $\nu_{v \rightarrow c}^{(0)} = \lambda_v$ to each check node $c \in \mathcal{N}(v)$.
At the $t^{\text{th}}$ iteration, $t \ge 1$, each check node $c$ passes the message $\mu_{c \rightarrow v}^{(t)}$ to each variable node $v \in \mathcal{N}(c)$, where
\begin{align}
\label{eq:bpchktovar}
\mu_{c \rightarrow v}^{(t)} = (-1)^{\un{s}_c} ~ 2 ~ \text{tanh}^{-1} \left( \prod_{v' \in \mathcal{N}(c)\setminus \{v \}} \text{tanh} \left( \frac{\nu_{v' \rightarrow c}^{(t-1)}}{2}\right) \right ).
\end{align}
Recall that $\un{s}_c$ denotes the $c$-th bit of $\un{s}$.
Subsequently, each variable node $v$ passes the message $\nu_{v \rightarrow c}^{(t)}$ to each check node $c \in \mathcal{N}(v)$, where
\begin{align}
\label{eq:bpvartochk}
\nu_{v \rightarrow c}^{(t)} = \lambda_v+\sum_{c' \in \mathcal{N}(v)\setminus \{ c\}} \mu_{c' \rightarrow v}^{(t)}.
\end{align}
Thus, apart from the $0^{\text{th}}$ iteration, each iteration of BP consists of two rounds of message-passing. In the first round, messages are passed from check nodes to variable nodes according to \eqref{eq:bpchktovar}, and in the second round, messages are passed from variable nodes to check nodes according to \eqref{eq:bpvartochk}.
After $T$ iterations, the final message at variable node $v$ is computed as
\begin{align*}
\gamma_{v} = \lambda_v+\sum_{c' \in \mathcal{N}(v)} \mu_{c' \rightarrow v}^{(T)}. 
\end{align*} 
The error estimate $\hat{\un{e}}$ is obtained by taking hard decisions on the variable nodes as
\begin{align*}
\hat{\un{e}}_v = \begin{cases}
    0, &\gamma_{v} > 0,\\
    1, &\gamma_{v} \leq 0.
\end{cases}
\end{align*}

If $H_Z\hat{\un{e}} =\un{s}$, we say that BP has converged, otherwise, we say that a non-convergence failure has occurred. Note that convergence only implies that $\un{e}$ and $\hat{\un{e}}$ have the same syndrome, i.e., the Pauli operator corresponding to $(\un{e}+\hat{\un{e}})$ belongs to the centralizer of the stabilizer group. Decoding is successful iff this Pauli operator belongs to the stabilizer group. Thus, if the decoder converges but the Pauli operator corresponding to $(\un{e}+\hat{\un{e}})$ does not belong to the stabilizer group, we have a logical error.

\subsection{Circuit-Level Noise}

When the stabilizers of the code are measured with a real syndrome extraction circuit, the effect of different noisy components adds up. In particular:
\begin{itemize}
    \item Each ancilla qubit is intended to be initialized in the $\ket{0}$ (or $\ket{+}$) state, but with probability $p$, the orthogonal state $\ket{1}$ (or $\ket{-}$) is prepared instead.
    \item The measurement of an ancilla qubit is flipped with probability $p$.
    \item Each single-qubit unitary gate is followed by a single-qubit depolarizing channel with error probability $p$. Similarly, every two-qubit gate (such as CNOT or CZ) is followed by a two-qubit depolarizing channel. In the latter case, with probability $p$, one of the 15 non-identity two-qubit Pauli operators $P \in \{I_2, X, Y, Z\}^{\otimes 2} \setminus \{I_2 \otimes I_2\}$ is applied.
    \item  Any qubit (data or ancilla) that is not involved in a gate during a time step undergoes a single-qubit depolarizing channel with probability $p$ (idling errors).
\end{itemize}

Circuit-level noise cannot be directly decoded using the original Tanner graph of the code for two main reasons: $i)$ the measurement of each stabilizer is repeated $d$ times, $ii)$ the decoder would fail to account for space-time correlated error mechanisms (hook errors), and $iii)$ the noise introduces syndrome outcomes that lie outside the column space of the original parity check matrix $H$. Consequently, decoding of circuit-level noise is carried out on a new Tanner graph called \textit{circuit-level Tanner graph}, or \textit{detector error model}, where each variable node corresponds to a fault mechanism, and each check node corresponds to a \textit{detector}, namely the parity between two subsequent measurements of the same stabilizer. For more details on the construction of the detector error model, see~\cite{derks2025designing,gong2024toward,borah2025fault}.

The code whose parity-check matrix is given by the circuit-level Tanner graph should be interpreted as a standalone quantum code. In the phenomenological noise setting, this object has been termed the \textit{outcome code}~\cite{hillmann_complexes}, and we adopt the same terminology for its circuit-level counterpart.
In~\cite{hillmann_complexes}, it was shown that under phenomenological noise the outcome code can be constructed as a hypergraph product between the underlying quantum code and a classical repetition code. An analogous characterization for circuit-level noise has not yet been established. Nevertheless, the circuit-level outcome code cannot be regarded as a purely classical code. Rather, it must be treated as a quantum stabilizer code, since it also exhibits harmless error patterns, commuting with all the check rows and the logical observables.
Therefore, although the following analysis is presented for a CSS code under the code-capacity noise model, the same considerations apply more broadly to the outcome code associated with circuit-level noise.

\section{Equivalence of Degeneracy and Code-Shortening}
\label{sec:degeneracy_and_shortening}
We now discuss the connection between degeneracy and code-shortening. In the following, without loss of generality, we consider only data errors, as a similar characterization holds for circuit-level noise, although we leave a more detailed analysis for future work. 

Consider an $[[n,k,d]]$ CSS code $\text{CSS}({\mathcal{C}_X,\mathcal{C}_Z})$, and let $H_X$ and $H_Z$ be parity-check matrices for \(\mathcal{C}_X\)
and \(\mathcal{C}_Z\), respectively. Recall that $H_X$ is assumed to be of the form $H_X=[I_\rho \mid A]$.
Consider an $X$-type error $\un{e}\in\mathbb{F}_2^n$ produced by the channel and let $\un{s}=H_Z\un{e}$ be the corresponding syndrome. Consider the index $i=1$. Clearly, either ${e}_1=0$ or ${e}_1=1$. If ${e}_1=1$, the degenerate error $\un{e}'=(\un{e}+(H_{X,\text{row},1})^T)$ satisfies ${e}'_1 = 0$. In fact, it is straightforward to see that any stabilizer that corresponds to a linear combination of the rows of $H_X$ that contains the first row can be added to $\un{e}$ to get a degenerate error of this form (and that all degenerate errors of this form are obtained in this way).
In conclusion, we are guaranteed that either the channel error $\un{e}$, or a degenerate error $\un{e}'$, has the property that its first entry is $0$. Thus, at the decoder, instead of trying to find a general $\un{\hat{e}}$ that satisfies $H_Z\un{\hat{e}}=\un{s}$ (which we will refer to as usual decoding henceforth), we may search only among those $\un{\hat{e}}$ which satisfy ${\hat{e}}_1=0$ (since we know that such an $\un{\hat{e}}$ exists). This leads us to the decoding problem:
\begin{align}
H_Z \hat{\un{e}} = \un{s}
&\iff \sum_{i=1}^{n} \hat{e}_i \, H_{Z,i} = \un{s} \nonumber \\
&\iff \hat{e}_1 H_{Z,1} + \sum_{i=2}^{n} \hat{e}_i \, H_{Z,i}
    = \un{s} \nonumber \\
&\iff 0\cdot H_{Z,1} + \sum_{i=2}^{n} \hat{e}_i \, H_{Z,i} = \un{s}.
\label{eq:shortening_0}
\end{align}

Now consider the code $\mathcal{C}_{Z,{\{1\},0}}$, the code obtained by shortening $\mathcal{C}_{Z}$  on the index $i=1$ to $0$. Recall that this is the collection of codewords of $\mathcal{C}_Z$ whose first coordinate is $0$. It is clear that the decoding problem in  \eqref{eq:shortening_0} can be cast as a decoding problem in $\mathcal{C}_{Z,{\{1\}},0}$, where the first bit is received error-free (or never transmitted, as is the case in practice, since in the classical case the decoder knows that a codeword from $\mathcal{C}_{Z,{\{1\}},0}$ is being transmitted), and the channel produces errors only on the bits indexed by $[2:n-1]$. Thus, degeneracy allows us to perform decoding on the shortened code $\mathcal{C}_{Z,{\{1\}},0}$ instead of the code $\mathcal{C}_{Z}$. Clearly, the same conclusion can be drawn for any index $i\in [n]$.

In the above discussion, we concluded, by exploiting degeneracy, that either the channel error $\un{e}$, or a degenerate error $\un{e}'$, has the property that its first entry is $0$. By the same reasoning, we can alternatively conclude that either the channel error $\un{e}$, or a degenerate error $\un{e}'$, has the property that its first entry is $1$.
In this case, the decoder searches for an error estimate $\un{\hat{e}}$ such that ${\hat{e}}_1=1$. Proceeding as before, we get 
\begin{align}
H_Z \hat{\un{e}} = \un{s} \iff 1\cdot H_{Z,1} + \sum_{i=2}^{n} \hat{e}_i \, H_{Z,i} = \un{s},
\label{eq:shortening_1}
\end{align}
analogous to \eqref{eq:shortening_0}. As before, \eqref{eq:shortening_1} can be cast as a decoding problem in ${C}_{Z,{\{1\}},1}$, the set of codewords of $\mathcal{C}_Z$ whose first coordinate is $1$. In other words, the decoder can perform decoding on the shortened code $\mathcal{C}_{Z,{\{1\}},1}$. Once again, a similar argument holds for any index $i\in [n]$.

It is noteworthy that degeneracy allows the decoder to shorten $\mathcal{C}_Z$ to both $0$ and $1$. This is because there exist degenerate errors with $0$ in the $i$-th coordinate, as well as degenerate errors with $1$ in the $i$-th coordinate, and the decoder can restrict its search to either of these subsets. 
We wish to highlight an important difference between the classical and the quantum case here. 
As mentioned before, the operation of shortening can not be performed at the decoder classically, since a classical decoder needs to identify the transmitted codeword (and hence the error produced by the channel) exactly. Suppose that the code $\mathcal{C}_Z$ is used for classical transmission. Unless $\mathcal{C}_Z$ is shortened by the transmitter, the transmitted codeword can take on value either $0$ or $1$ in the $i$-th coordinate, and the classical decoder can not restrict itself to performing decoding only on ${C}_{Z,{\{1\}},0}$ or only on ${C}_{Z,{\{1\}},1}$. However, in the quantum case, due to degeneracy, restricting its attention to either ${C}_{Z,{\{1\}},0}$ or ${C}_{Z,{\{1\}},1}$ does not impair the decoder's ability to find a valid error estimate. 

We now turn to a comparison and analysis of the two cases considered in \eqref{eq:shortening_0} and \eqref{eq:shortening_1}, respectively.
Before discussing their differences, we first highlight a connection between the two. We will later use similar ideas to provide an alternative way to arrive at the definition of the minimum distance of a CSS code provided in \eqref{eq:dmin_of_CSS}. We proceed from \eqref{eq:shortening_1},
\begin{align}
H_Z \hat{\un{e}} = \un{s} &\iff 1\cdot H_{Z,1} + \sum_{i=2}^{n} \hat{e}_i \, H_{Z,i} = \un{s}, \nonumber\\
&\iff \sum_{i=2}^{n} \hat{e}_i \, H_{Z,i}
    = \un{s} - \hat{e}_1 H_{Z,1}
    = \un{s}' \nonumber \\
&\iff 0\cdot H_{Z,1} + \sum_{i=2}^{n} \hat{e}_i \, H_{Z,i} = \un{s}'. \nonumber
\end{align}
It follows from that a decoding problem in ${C}_{Z,{\{1\}},1}$ can be translated into a decoding problem in ${C}_{Z,{\{1\}},0}$, simply by modifying the syndrome appropriately. This is not surprising since, as mentioned before, ${C}_{Z,{\{1\}},1}$ is a coset of ${C}_{Z,{\{1\}},0}$ in $\mathcal{C}_Z$. Thus, ${C}_{Z,{\{1\}},1}$ may be regarded as a deterministic translate of ${C}_{Z,{\{1\}},0}$, and thus decoding in ${C}_{Z,{\{1\}},1}$ is essentially decoding in ${C}_{Z,{\{1\}},0}$.

We now examine some qualitative differences among the three scenarios: $(i)$ usual decoding, $(ii)$ shortening to $0$ as in \eqref{eq:shortening_0}, and $(iii)$ shortening to $1$ as in \eqref{eq:shortening_1}. The discussion is in the context of iterative decoding and under the assumption that the physical error rate $p$ is small.
Firstly, by restricting attention to either only ${C}_{Z,{\{1\}},0}$ or ${C}_{Z,{\{1}\},1}$, the number of degenerate errors within the decoder’s search space in $(ii)$ and $(iii)$ is reduced by a factor of two compared to $(i)$; moreover, the search space itself that the decoder needs to explore in $(ii)$ or $(iii)$ is half of that in $(i)$.
Consequently, iterative decoding is more likely to succeed in $(ii)$ and $(iii)$ than in $(i)$. Next,
since the physical error rate $p$ is small, the probability that the error $\un{e}$ produced by the channel is such that ${e}_1=1$ is small,
and hence shortening $\mathcal{C}_Z$ on index $i=1$ to $0$ typically does not force the decoder to search for a degenerate error. On the contrary, it correctly identifies the value of ${e}_1$ with probability $(1-p)$. One may qualify this situation as follows: the decoders in $(i)$ and $(ii)$ try to estimate the same error $\un{e}$, where the decoder in $(ii)$ guesses the first bit of $\un{e}$ correctly with large probability. On the other hand, when $p$ is small, shortening to $1$ as in \eqref{eq:shortening_1} typically forces the decoder to search for a degenerate error. One may expect this to be detrimental, since the weight of a degenerate error may be larger than that of the original error.  Consider for example the extreme case when the channel doesn't produce any errors, i.e., the case when $\un{e}=\un{0}$. Clearly, degenerate errors $\un{e}'$ satisfying ${e}'_1=1$ correspond to non-identity $X$-type stabilizers whose support contains the index $i=1$. The weight of any such stabilizer is obviously larger than $0$. However, simulation results presented in Section \ref{sec:Numerical_Results} show that shortening to $1$ clearly outperforms shortening to $0$ (for example, see Fig. \ref{fig:BB288_0_1}). We shall provide an explanation for this behavior and present several related observations from our simulations in Section \ref{sec:Numerical_Results}. 

We end this section with a discussion of how the code-shortening perspective provides a simple understanding of the distance of a CSS code as defined in \eqref{eq:dmin_of_CSS}. In the following, we will refer to $\un{v}_{[\rho]}=(v_1,v_2,\dots,v_\rho)^T$ as the prefix of the vector $\un{v} = (v_1,v_2,\dots,v_n)^T \in \mathbb{F}^n_2$. Consider an error pattern $\un{e}$ produced by the channel. Then, since $H_X$ has the form $H_X=[I_\rho\mid A]$, for any $\un{b}\in \mathbb{F}_2^\rho$, there exists a (possibly) degenerate error $\un{e}'$ such that $\un{e}'_{[\rho]}=\un{b}$. Further, since $I_\rho$ has full rank, $\un{e}'$ is the unique degenerate error with prefix $\un{b}$. Now, suppose that a genie provides us the prefix $\un{b}^*$ of a least weight degenerate error (but not the full error vector, since then there would be nothing left to decode). We can then set up the following decoder that searches for an error estimate among only those vectors that have prefix $\un{b}^*$:
\begin{align*}
H_Z \hat{\un{e}} = \un{s} &\iff \sum_{i=1}^{\rho} b^*_i \, H_{Z,i} + \sum_{i=\rho + 1}^{n} \hat{e}_i \, H_{Z,i} = \un{s}, \\
&\iff \sum_{i=\rho + 1}^{n} \hat{e}_i \, H_{Z,i}
    = \un{s} - \sum_{i=1}^{\rho} b^*_i \, H_{Z,i}
    = \un{s}' \\
&\iff \sum_{i=1}^{\rho}0\cdot H_{Z,i} + \sum_{i=\rho + 1}^{n} \hat{e}_i \, H_{Z,i} = \un{s}'.
\end{align*}
We thus end up in a (classical) decoding problem in $\mathcal{C}_{Z,[\rho]}$, the code $\mathcal{C}_Z$ shortened on the coordinates in $[\rho]$. Note that at this point, there is a unique correct answer that the decoder must search for (the effect of degeneracy has been completely absorbed into the knowledge of $\un{b}^*$, or equivalently, the modified syndrome). Recall that $G_Z$ is of the form \eqref{eq:Gz_form}. It follows that codewords of $\mathcal{C}_{Z,[\rho]}$, i.e., codewords of $C_Z$ that have prefix $\un{0}_\rho$, are precisely those that are obtained by taking linear combinations of the rows of $L_X$. Thus, $L_X$, and analogously $L_Z$ (for the case of $Z$-type errors), determine the error-correcting capability of the stabilizer code, which is precisely what is captured by the definition of minimum distance in \eqref{eq:dmin_of_CSS}.

\section{Impulse Decoding: A Parallel Decoding Scheme for QLDPC Codes}
\label{sec:Impulse_Decoder}
In this section, we present a parallel decoding scheme where each decoder shortens a different variable node (here we use the phrase shortening a variable node to refer to shortening performed on the index corresponding to the variable node). We refer to this scheme as \emph{impulse decoding}, since in the setting of BP decoding, shortening can be realized by assigning infinite LLRs to variable nodes. In particular, setting the initial LLR of a variable node $v$ to $\lambda_v = +\infty$ corresponds to shortening to $0$, while setting $\lambda_v = -\infty$ corresponds to shortening to $1$.
 
The pseudocode for impulse decoding is presented in Algorithm~\ref{algo:impulse_decoder}. Consider an $[[n,k,d]]$ CSS QLDPC code and let $\mathcal{T}=(\mathcal{V}, \mathcal{C}, \mathcal{E})$ denote the Tanner graph corresponding to $H_Z$. We begin by performing usual BP decoding first. In the event that the decoder does not converge, we proceed by shortening variable nodes. Specifically, we employ $n$ parallel decoders (we will see later that the number of parallel decoders required in practice is much lesser), labeled $\mathcal{D}_1,\dots,\mathcal{D}_n$, where decoder $\mathcal{D}_i$ shortens the $i$-th variable node to $1$ by setting its initial LLR to $-\infty$ (we to shorten to $1$ since this yields better performance than shortening to 0, see Section \ref{sec:Numerical_Results}). With only this simple modification in the initial LLRs, decoder $\mathcal{D}_i$ proceeds with BP decoding. We remark here that some parallel decoders in the literature (e.g., \cite{koutsioumpas2025automorphism}) require a modification of the Tanner graph itself for different parallel branches, while initializing the parallel decoders in our scheme is straightforward. We further remark that, unlike many schemes in the literature, we do not use the initial run of BP to perform post-processing. 
In principle, it could also have been performed in parallel with the decoders $\mathcal{D}_1,\dots,\mathcal{D}_n$. However, since BP is a powerful algorithm that converges for a large number of instances, invoking the decoders $\mathcal{D}_1, \dots, \mathcal{D}_n$ in every case is unnecessarily excessive.
Further, employing the parallel decoders only when necessary reduces hardware power consumption, and it is standard practice in the literature to resort to alternative techniques only when BP fails. 
We will later discuss a natural approach that uses the output of the initial BP run to make a more informed selection of candidate variable nodes for shortening when the number of parallel decoders available is much smaller than the number of variable nodes in the Tanner graph (as may occur, for example, in circuit-level Tanner graphs). In this setting, our scheme may also be viewed as a form of post-processing, however, as we will discuss later, our approach incurs very little additional complexity. We wish to emphasize that while the idea of using BP outputs to fix the values of certain variable nodes has gained traction in the recent literature \cite{valentini2025restartbeliefgeneralquantum,ye2025beamsearchdecoderquantum,muller2025improved,gong2024toward,yao2024belief,alinia2025decimation,tsubouchi2025degeneracy}, it has largely been motivated by heuristic arguments and empirical observations. By making the connection to code-shortening, we provide a more principled basis for this approach. Moreover, as shown by simulation results in the next section, for short codes such as the popular $[[288,12,18]]$ and $[[144,12,12]]$ BB codes, it may not be necessary to incur the complexity of processing BP outputs when a reasonable number of parallel decoders is available.

\begin{algorithm}[t]
\caption{Impulse Decoding}
\textbf{Input:} Tanner graph $\mathcal{T}(H)=(\mathcal{V},\mathcal{C},\mathcal{E})$, syndrome $s$\\
\textbf{Output:} Error estimate $\hat{e}$ or non-convergence
\hrule
\begin{algorithmic}[1]

\STATE $\lambda \gets \ln\!\left(\frac{1-p}{p}\right)$
\STATE Initialize channel LLR vector $\Lambda = (\lambda,\dots,\lambda)$

\STATE $(\hat{\un{e}}, \text{flag}_0) \gets \mathrm{BP}(\mathcal{T}(H),\Lambda,s)$

\IF{$\text{flag}_0 = \text{converged}$}
    \STATE \textbf{return} $\hat{\un{e}}$
\ENDIF

\ParFor{$i = 1$ to $n$}
    \STATE $\Lambda^{(i)} \gets \Lambda$
    \STATE $\Lambda^{(i)}_i \gets -\infty$ \COMMENT{shorten variable node $i$ to $1$}
    \STATE $(\hat{\un{e}}^{(i)},\text{flag}_i) \gets \mathrm{BP}(\mathcal{T}(H),\Lambda^{(i)},s)$
\EndParFor

\STATE $\mathcal{I} \gets \{ i \in [1:n] \mid \text{flag}_i = \text{converged} \}$

\IF{$\mathcal{I} = \emptyset$}
    \STATE \textbf{return} non-convergence
\ENDIF

\STATE $\hat{\un{e}} \gets \arg\min_{i\in\mathcal{I}} \mathrm{wt}(\hat{\un{e}}^{(i)})$

\STATE \textbf{return} $\hat{\un{e}}$

\end{algorithmic}
\label{algo:impulse_decoder}
\end{algorithm}

We return to the description of Algorithm \ref{algo:impulse_decoder}. If the initial BP decoding converges, the error estimate given by it is considered as the final error estimate. In case the initial BP decoding does not converge, we employ the parallel decoders $\mathcal{D}_1,\dots,\mathcal{D}_n$, such that decoder $\mathcal{D}_i$ yields the error estimate $\hat{\un{e}}^{(i)}$. Let $\mathcal{I} \subseteq [1:n]$ be the subset of indices $i$ such that $\mathcal{D}_i$ converges. The final error estimate is obtained by selecting the minimum-weight error estimate among all converged decoders, i.e., 
$$
    \hat{\un{e}} = \arg\min_{ i\in\mathcal{I}}\ \text{wt}(\hat{\un{e}}^{(i)}).
$$
We shall refer to this criterion by the label $\mathtt{Minimum\_Weight}$.
An alternative criterion to select the final error estimate is to simply choose the error estimate corresponding to the smallest index $i^*$ in $\mathcal{I}$, i.e.,
\begin{equation*}
    \hat{\un{e}} = \hat{\un{e}}^{(i^*)}, \ \text{where}\ i^* = \min_{ i\in\mathcal{I}} i.
\end{equation*}
We shall refer to this criterion by the label $\mathtt{First\_Convergence}$. This criterion has two advantages over the $\mathtt{Minimum\_Weight}$ criterion. First, it avoids the additional latency and complexity incurred in finding the least weight error estimate among the decoders indexed by $\mathcal{I}$. Second, for a serial implementation of Algorithm \ref{algo:impulse_decoder} where variable nodes are shortened sequentially, the $\mathtt{First\_Convergence}$ criterion significantly improves decoding latency (we will discuss more about this in the next section). However, this comes at the cost of an increased logical error rate, especially at lower physical error rates (for example, see Fig. \ref{fig:BB288_0_1}).

\subsection{Impulse Decoding for Circuit-Level Noise}

We now discuss the case when the Tanner graph has a large number of variable nodes. This is typical in circuit-level Tanner graphs. For example, the circuit-level Tanner graph of the $[[90,8,10]]$ BB code 
considered in the next section has approximately $4000$ variable nodes, while that of the $[[144,12,12]]$ BB code has approximately $8000$ variable nodes \cite{gong2024toward}. Clearly, shortening each variable node is not practical, making it necessary to incur some additional complexity in selecting candidate variable nodes for shortening.

We adopt the natural approach of selecting the least reliable variable nodes for shortening, based on their reliabilities obtained from the initial BP run. The reliability of variable node $v$ is the absolute value $|\gamma_v|$ of the final message at that variable node.
Specifically, suppose that $N$ parallel decoders are available. We first perform an initial BP run, and if it fails to converge, we activate the parallel decoders $\mathcal{D}_1,\dots,\mathcal{D}_N$, where decoder $\mathcal{D}_i$ shortens the $i$-th least reliable variable node. If one or more decoders converge, we select the final error estimate based on the $\mathtt{Minimum\_Weight}$ criterion and do not proceed to further rounds. However, if none of the decoders converge in the first round, we proceed to shorten the next $N$ least reliable variable nodes. This process continues until either convergence is achieved or a maximum number $R$ of rounds is reached, at which point we declare non-convergence. The pseudocode for this procedure is provided in Algorithm \ref{algo:reliability_impulse_decoder}. Note that once again, to improve latency, the $\mathtt{First\_Convergenece}$ criterion may be employed in place of the $\mathtt{Minimum\_Weight}$ criterion to choose the final error estimate.

\begin{algorithm}
\caption{Reliability-Based Impulse Decoding}
\textbf{Input:} Tanner graph $\mathcal{T}(H)=(\mathcal{V},\mathcal{C},\mathcal{E})$, channel LLR vector $\Lambda$, syndrome $s$, number of parallel decoders $N$, maximum number of rounds $R$\\
\textbf{Output:} Error estimate $\hat{\un{e}}$ or non-convergence
\hrule
\begin{algorithmic}[1]

\STATE $(\hat{\un{e}}, \text{flag}_0, \Lambda^{\mathrm{final}}) \gets \mathrm{BP}(\mathcal{T}(H),\Lambda,s)$

\IF{$\text{flag}_0 = \text{converged}$}
    \STATE \textbf{return} $\hat{\un{e}}$
\ENDIF

\STATE $\pi \gets \mathrm{argsort}(|\Lambda^{\mathrm{final}}|)$ \COMMENT{indices in increasing order of reliability}

\FOR{$r = 1$ to $R$}
    \STATE $\mathcal{J}_r \gets \{ \pi_{(r-1)N+1}, \dots, \pi_{rN} \}$

    \ParFor{$i \in \mathcal{J}_r$}
        \STATE $\Lambda^{(i)} \gets \Lambda$
        \STATE $\Lambda^{(i)}_i \gets -\infty$ \COMMENT{shorten variable node $i$ to $1$}
        \STATE $(\hat{\un{e}}^{(i)}, \text{flag}_i) \gets \mathrm{BP}(\mathcal{T}(H),\Lambda^{(i)},s)$
    \EndParFor

    \STATE $\mathcal{I}_r \gets \{ i \in \mathcal{J}_r \mid \text{flag}_i = \text{converged} \}$

    \IF{$\mathcal{I}_r \neq \emptyset$}
        \STATE $\hat{\un{e}} \gets \arg\min_{\hat{\un{e}}^{(i)}:\, i\in \mathcal{I}_r} \mathrm{wt}(\hat{\un{e}}^{(i)})$
        \STATE \textbf{return} $\hat{\un{e}}$
    \ENDIF

\ENDFOR

\STATE \textbf{return} non-convergence

\end{algorithmic}
\label{algo:reliability_impulse_decoder}
\end{algorithm}

We now discuss the complexity associated with Algorithm \ref{algo:impulse_decoder} and Algorithm \ref{algo:reliability_impulse_decoder}. We first focus on Algorithm \ref{algo:reliability_impulse_decoder}. As mentioned before, setting up the parallel decoders is straightforward and does not incur any complexity. The complexity associated with performing $T$ message-passing iterations of BP on a single decoder is $O(n T)$, where $n$ is the number of variable nodes in the Tanner graph and $\log(n)$ is a typical choice for $T$ \cite{dembo2010ising}. The complexity associated with sorting variable nodes according to their reliabilities is $O(n \log(n))$. Since in the worst case the algorithm performs $(R+1)$ rounds of decoding—an initial BP round followed by up to $R$ rounds, each using $N$ parallel decoders—and selects the final error estimate according to the $\mathtt{Minimum\_Weight}$ criterion, the total work complexity of Algorithm \ref{algo:reliability_impulse_decoder} is $O(R N n\log(n))$, while its parallel complexity is $O(R n\log(n))$. Algorithm \ref{algo:impulse_decoder} can be viewed as a special case of Algorithm \ref{algo:reliability_impulse_decoder}, with $R=1$ and $N=n$.

\subsection{Residual-Error-Based Impulse Decoding}
We now present another algorithm for decoding in the presence of circuit-level noise which requires (in practice) fewer parallel decoders than Algorithm \ref{algo:reliability_impulse_decoder}, but performs serial processing on each decoder.  The key idea behind this algorithm is the following. Consider the case that the channel produced the error $\un{e}$ yielding syndrome $\un{s} = H \un{e}$; the decoder then attempts to solve the system of equations $H \hat{\un{e}} = \un{s}$. Suppose that the decoder does not converge. Then, we are left with an error estimate $\hat{\un{e}}'$, with corresponding syndrome $ \un{s}' = H \hat{\un{e}}'$. We now proceed to decode for the residual error $\hat{\un{e}}'' = ({\un{e}} + \hat{\un{e}}')$, using the syndrome $\un{s}''=\un{s}+\un{s}'$ (since we know both these terms).
Clearly, if we succeed in finding such an $\hat{\un{e}}''$, then $\hat{\un{e}} = (\hat{\un{e}}' + \hat{\un{e}}'')$ satisfies $H \hat{\un{e}} = \un{s}$, as desired. On the other hand, if the decoder does not converge to such an $\hat{\un{e}}''$, we are left with an error estimate $\hat{\un{e}}'''$ with corresponding syndrome $H\hat{\un{e}}'''=\un{s}'''$, and we can attempt to solve for the residual error $(\hat{\un{e}}+\hat{\un{e}}''')$ using the syndrome $(\un{s}+\un{s}''')$. We can alternatively also try to solve for the error $({\un{e}}+\hat{\un{e}}'+\hat{\un{e}}''')$ using the syndrome $(\un{s}+\un{s}'+\un{s}''')$, but in this case we will have to store both $\hat{\un{e}}'$ and $\hat{\un{e}}'''$ in order to obtain the desired $\hat{\un{e}}$, as opposed to storing only $\hat{\un{e}}'''$ as above. 

In Algorithm \ref{algo:residual_error}, we combine this residual error decoding approach with impulse decoding performed by the $N$ parallel decoders in Algorithm \ref{algo:reliability_impulse_decoder}. In particular, the first round, which consists of parallelly shortening the $N$ least reliable variable nodes, is common to both algorithms. However, in the subsequent rounds of Algorithm \ref{algo:residual_error}, each decoder attempts to decode the residual error as described above, while keeping the corresponding variable node shortened throughout the decoding process. If any of the $N$ parallel decoders converges in the first round, the remaining rounds are skipped. Otherwise, each decoder continues until either convergence is achieved or a maximum number of rounds $R$ is reached. Finally, among all converged decoders, the least-weight error estimate is selected as the final output.

The complexity of Algorithm \ref{algo:residual_error} is similar to that of Algorithm \ref{algo:reliability_impulse_decoder}, though it is worth mentioning that while Algorithm \ref{algo:reliability_impulse_decoder} is highly parallelizable, the residual error decoding rounds in Algorithm \ref{algo:residual_error} need to be performed serially, which may increase latency. However, simulation results presented in Section \ref{sec:Numerical_Results} demonstrate that Algorithm \ref{algo:residual_error} outperforms Algorithm \ref{algo:reliability_impulse_decoder}, even while employing a much smaller number of parallel decoders and using only a few rounds (e.g., $N=20, R=6$).

\begin{algorithm}
\caption{Residual-Error-Based Impulse Decoding}
\textbf{Input:} Tanner graph  $\mathcal{T}(H)=(\mathcal{V},\mathcal{C},\mathcal{E})$, channel LLR vector $\Lambda$, syndrome $s$, number of parallel decoders $N$, maximum number of rounds $R$\\
\textbf{Output:} Error estimate $\hat{\un{e}}$ or non-convergence
\hrule
\begin{algorithmic}[1]

\STATE $(\hat{\un{e}},\Lambda^{\mathrm{final}}) \gets \mathrm{BP}(\mathcal{T}(H),\Lambda,s)$

\IF{$H\hat{\un{e}} = s$}
    \STATE \textbf{return} $\hat{\un{e}}$
\ENDIF

\STATE $\pi \gets \mathrm{argsort}(|\Lambda^{\mathrm{final}}|)$
\COMMENT{indices in increasing order of reliability}
\STATE $\mathcal{L} \gets \emptyset$

\ParFor{$j = 1$ to $N$}
    \STATE $i \gets \pi_j$
    \STATE $\Lambda^{(i)} \gets \Lambda$
    \STATE $\Lambda^{(i)}_i \gets -\infty$
    \COMMENT{shorten variable node $i$ to $1$}

    \STATE $(\hat{\un{e}}^{(i)},-) \gets \mathrm{BP}(\mathcal{T}(H),\Lambda^{(i)},s)$

    \IF{$H\hat{\un{e}}^{(i)} = s$}
        \STATE $\mathcal{L} \gets \mathcal{L} \cup \{\hat{\un{e}}^{(i)}\}$
    \ENDIF

    \STATE $\hat{\un{e}}_{\mathrm{base}}^{(i)} \gets \hat{\un{e}}^{(i)}$
\EndParFor

\IF{$\mathcal{L} \neq \emptyset$}
    \STATE $\hat{\un{e}} \gets
    \arg\min_{\hat{\un{e}}'\in\mathcal{L}} \mathrm{wt}(\hat{\un{e}}')$
    \STATE \textbf{return} $\hat{\un{e}}$
\ENDIF

\ParFor{$j = 1$ to $N$}
    \STATE $i \gets \pi_j$

    \FOR{$r = 2$ to $R$}

        \STATE $s_{\mathrm{res}}^{(i)} \gets s \oplus H\hat{\un{e}}_{\mathrm{base}}^{(i)}$

        \STATE $(\hat{\un{e}}_{\mathrm{res}}^{(i)},-) \gets
        \mathrm{BP}(\mathcal{T}(H),\Lambda^{(i)},s_{\mathrm{res}}^{(i)})$

        \IF{$H\hat{\un{e}}_{\mathrm{res}}^{(i)} = s_{\mathrm{res}}^{(i)}$}
            \STATE $\hat{\un{e}}^{(i)} \gets
            \hat{\un{e}}_{\mathrm{base}}^{(i)} \oplus \hat{\un{e}}_{\mathrm{res}}^{(i)}$
            \STATE $\mathcal{L} \gets \mathcal{L} \cup \{\hat{\un{e}}^{(i)}\}$
            \STATE \textbf{break}
        \ENDIF

        \STATE $\hat{\un{e}}_{\mathrm{base}}^{(i)} \gets \hat{\un{e}}_{\mathrm{res}}^{(i)}$

    \ENDFOR
\EndParFor

\IF{$\mathcal{L} \neq \emptyset$}
    \STATE $\hat{\un{e}} \gets
    \arg\min_{\hat{\un{e}}'\in\mathcal{L}} \mathrm{wt}(\hat{\un{e}}')$
    \STATE \textbf{return} $\hat{\un{e}}$
\ENDIF

\STATE \textbf{return} non-convergence

\end{algorithmic}
\label{algo:residual_error}
\end{algorithm}

\section{Numerical Results}
\label{sec:Numerical_Results}

We now present simulation results\footnote{All relevant simulation details are provided to ensure reproducibility. The simulation data are available from the authors upon reasonable request.}. We first focus on Algorithm \ref{algo:impulse_decoder}. For all corresponding plots, $T=100$ message-passing iterations are performed each time BP is performed. Further, we limit the values of the messages passed from variable to check nodes to the range $[-25, 25]$ (this is commonly done for numerical stability, see for example \cite{yao2024belief,alinia2025decimation}).  Failure rate in the plots indicates failures due to both non-convergence and logical errors. All plots are obtained by collecting at least 100 failures. The plots for BP-OSD follow \cite{roffe2020decoding,gong2024toward,yao2024belief}. Throughout this section, including the figures, we use the label $\mathtt{Minimum\_Weight}/0$ to refer to the setting where variable nodes are shortened to 0, and the $\mathtt{Minimum\_Weight}$ criterion is used to choose the final error estimate. Similarly, we use the labels $\mathtt{Minimum\_Weight}/1$, $\mathtt{First\_Convergence}/0$ and $\mathtt{First\_Convergence}/1$.

\begin{figure}[ht]
    \centering
    \includegraphics[width=0.9\linewidth]{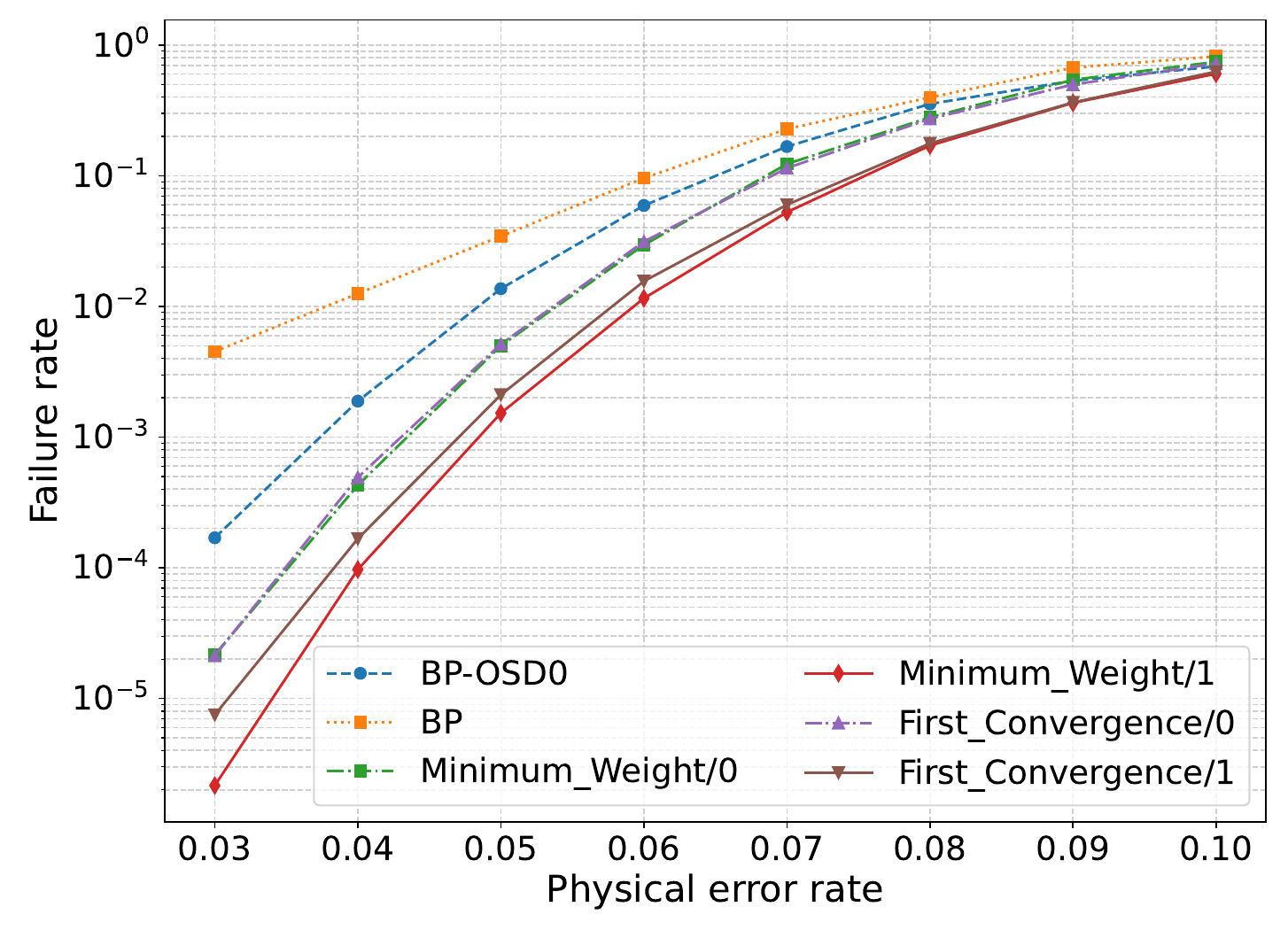}
    \caption{Performance of impulse decoding (Algorithm \ref{algo:impulse_decoder}) for the $[[288,12,18]]$ BB code. 
    The labels $0$ and $1$ in the legend indicate shortening to $0$ and $1$, respectively.
    }
    \label{fig:BB288_0_1}
\end{figure}

Fig. \ref{fig:BB288_0_1} shows the performance of Algorithm \ref{algo:impulse_decoder} for the $[[288,12,18]]$ BB code. It can be clearly seen that impulse decoding signigicantly outperforms BP-OSD0, especially at lower error rates. The figure also shows that shortening to $1$ achieves notably better performance than shortening to $0$. As discussed in Section \ref{sec:Impulse_Decoder}, since physical error rates are low (of the order of $10^{-2}$), the channel error $\un{e}$ has small weight. When shortening variable nodes to $0$, each of the parallel decoders corresponding to the variable nodes in the support of $\un{e}$ is forced to search for a degenerate error, while all the remaining decoders search for $\un{e}$ (typically, since all other degenerate errors are likely to have larger weight).
The situation for shortening to $1$ is complimentary, with all decoders corresponding to variable nodes not in the support of $\un{e}$ searching for a degenerate error. Thus, shortening to $1$ enables a much broader exploration of the space of degenerate errors. Fig. \ref{fig:BB288_0_1} shows that this strategy is indeed effective, suggesting that encouraging convergence to a degenerate error can be more beneficial than the conventional approach of seeking the channel-induced error itself. Several observations presented below further support this principle, providing strong evidence that shortening to $1$ is particularly favorable for decoding, both in terms of performance and latency.

We now discuss and compare the various curves in Fig. \ref{fig:BB288_0_1}.
Observe that for moderate and high error rates, the performance gap between the $\mathtt{Minimum\_Weight}$ and $\mathtt{First\_Convergence}$ criteria is not that significant when shortening to 1, however, in the low error-rate regime, the $\mathtt{Minimum\_Weight}$ criterion yields noticeably better performance. On the other hand, when shortening to 0, there is virtually no difference between the two criteria across all physical error rates. This behavior can be explained with the help of Table \ref{tab:shortening_trend}, where we record the percentages of degenerate and logical errors corresponding to the various curves in Fig. \ref{fig:BB288_0_1} (where we have subtracted the degenerate and logical errors that occur due to the initial BP run, so that only the effect of shortening is captured). Several interesting trends can be observed in Table \ref{tab:shortening_trend}. First, there is a striking difference between shortening to 0 and shortening to 1, under both the $\mathtt{Minimum\_Weight}$ and $\mathtt{First\_Convergence}$ criteria, in the percentage of logical errors -- shortening to 1 leads to a significantly larger percentage of logical errors than shortening to 0. Further, as the physical error rate decreases, the percentage of logical as well as degenerate errors increases sharply under $\mathtt{First\_Convergence}/1$. This shows that shortening to 1 is highly effective at making the decoder converge, and while this convergence is often to a degenerate error, there is also a higher likelihood of converging to a logical error. Selecting the minimum weight error estimate therefore helps in decreasing the probability of making a logical error. Note that for $\mathtt{Minimum\_Weight}/1$, while the percentage of logical errors increases (though not as sharply as $\mathtt{First\_Convergence}/1$) with decreasing physical error rate, the percentage of degenerate errors stays approximately constant, which is closer to the behavior exhibited by $\mathtt{First\_Convergence}/0$ and $\mathtt{Minimum\_Weight}/0$. Also observe that the number of logical errors under $\mathtt{First\_Convergence}/0$ is very small, which explains why the performance of $\mathtt{Minimum\_Weight}/0$ is almost identical to that of $\mathtt{First\_Convergence}/0$.

\begin{table}[t]
\centering
\caption{Percentages of degenerate and logical errors observed under various settings. The pair $(a,b)$ denotes that out of the total number of successes, $a\%$ were degenerate errors, and out of the total number of failures, $b\%$ were logical errors.}
\begin{tabular}{c|ccc}
\hline
 & $p=0.03$ & $p=0.04$ & $p=0.05$ \\
\hline
$\mathtt{First\_Convergence}/0$ & (55.21, 4) & (54.97, 4.2) & (49.77, 4.4) \\ 
$\mathtt{First\_Convergence}/1$ & (93.4, 80) & (88, 69.8) & (75.89, 59.4) \\
$\mathtt{Minimum\_Weight}/0$ & (54.71, 6) & (54.52, 4) & (49.30, 3.8) \\
$\mathtt{Minimum\_Weight}/1$ & (54.3, 50) & (56.67, 48.2) & (55.03, 38.20) \\
\hline
\end{tabular}
\label{tab:shortening_trend}
\end{table}

\begin{figure}[htbp]
    \centering

    \begin{subfigure}{0.48\textwidth}
        \centering
        \includegraphics[width=\linewidth]{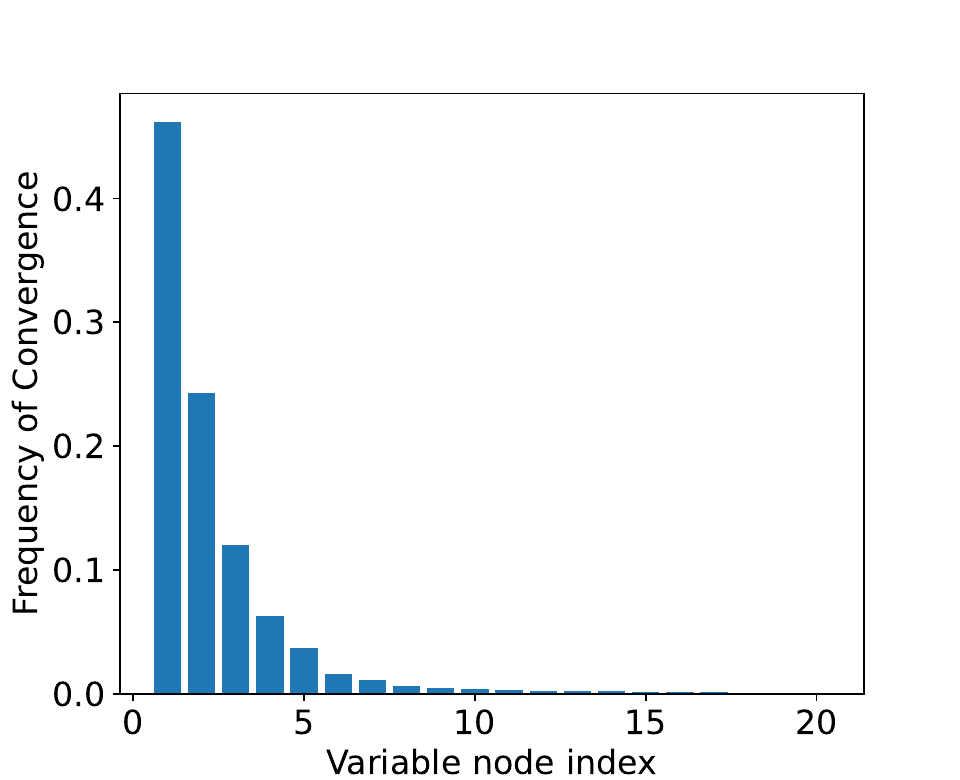}\caption{$\mathtt{First\_Convergence/1}$.}
        \label{fig:BB288_hist_0.03}
    \end{subfigure}
    \hfill
    \begin{subfigure}{0.48\textwidth}
        \centering
        \includegraphics[width=\linewidth]{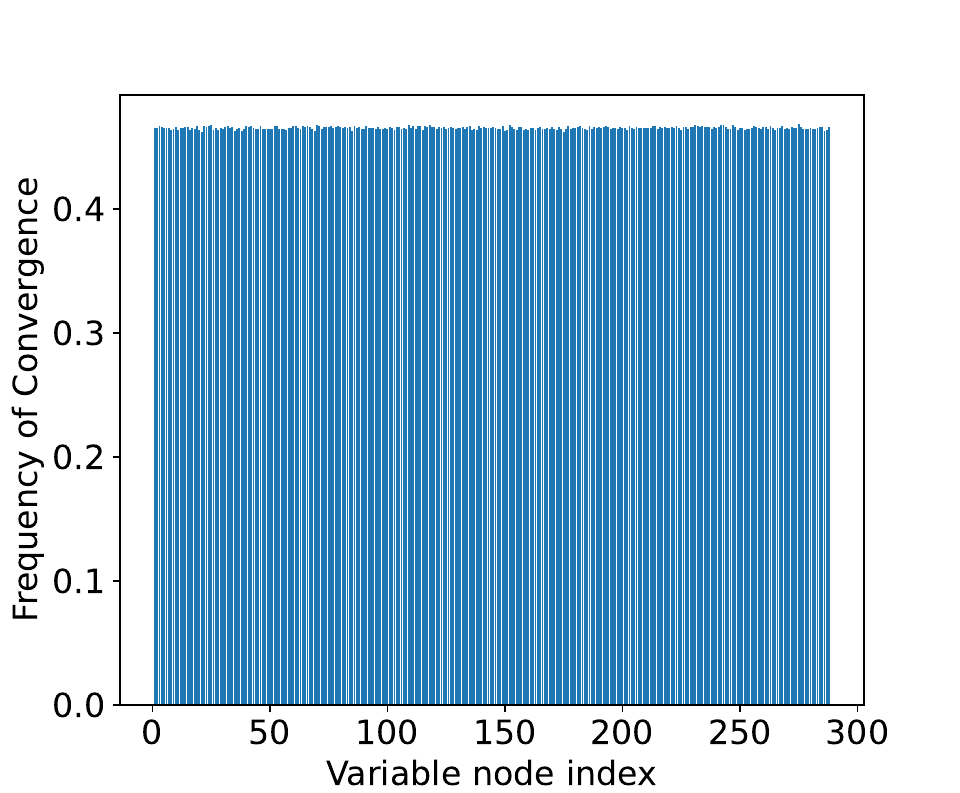}
        \caption{$\mathtt{Minimum\_Weight}/1$.}
        \label{fig:BB288_hist_min}
    \end{subfigure}

    \caption{Frequency of convergence upon shortening variable nodes under (a) $\mathtt{First\_Convergence/1}$ and (b) $\mathtt{Minimum\_Weight}/1$, for the $[[288,12,18]]$ BB code at physical error rate $0.03$.} 
    \label{fig:BB_288_latency}
\end{figure}

With regard to decoding latency, the $\mathtt{First\_Convergence}$ criterion offers a significant improvement over the $\mathtt{Minimum\_Weight}$ criterion. The latency under $\mathtt{First\_Convergence/1}$ for physical error rate $0.03$ is illustrated in Fig. \ref{fig:BB288_hist_0.03}, where the X-axis corresponds to variable node index, and the Y-axis corresponds to the frequency of convergence upon shortening variable node $i$ (the number of times convergence occurred upon shortening variable node $i$ normalized by the number of times shortening was employed, i.e., the number of times the initial round of BP decoding in Algorithm \ref{algo:impulse_decoder} failed), where, in accordance with the $\mathtt{First\_Convergence}$ criterion, variable nodes are shortened sequentially, with variable node $i$ being shortened only if shortening all preceding variable nodes failed to produce convergence. Thus, smaller values on the Y-axis for larger indices on the X-axis indicate low average latency. Fig. \ref{fig:BB288_hist_0.03} shows that the latency drops exponentially with the number of variable nodes shortened. This is interesting since the channel produces errors randomly, whereas shortening is performed sequentially. Indeed, on average, each variable node is equally likely to result in convergence upon being shortened, as illustrated by Fig. \ref{fig:BB288_hist_min}, which is the analogue of Fig. \ref{fig:BB288_hist_0.03} for the $\mathtt{Minimum\_Weight}/1$ setting (we will contrast this behavior with Fig. \ref{fig:B2_latency} later, where some variable nodes are more likely to yield convergence upon shortening than others).
Fig. \ref{fig:BB_288_latency} again provides a strong indication that decoder dynamics, complemented by graph connectivity, make shortening to 1 favorable for decoder convergence. 

Moreover, we observe that the number of BP iterations required for convergence also decreases upon shortening to 1. In particular, if we perform $T' = 50$ message-passing iterations in the parallel decoders in Algorithm \ref{algo:impulse_decoder},
the performance of $\mathtt{Minimum\_Weight}/1$ and $\mathtt{First\_Convergence/1}$ remains virtually the same. On the other hand, performance degrades when shortening is to 0 and $T' = 50$ message-passing iterations are performed.

Interestingly, note that in many instances where the decoder does not converge upon shortening a variable node to $1$, the residual error is supported only on the shortened node. That is, flipping the bit corresponding to the shortened node from $1$ to $0$ in the error estimate produced by BP yields an error estimate that satisfies the desired syndrome equation. Incorporating this additional stopping criterion significantly reduces the decoding latency (under the $\mathtt{First\_Convergence}$ criterion). However, somewhat surprisingly, it does not improve the overall decoding performance. We observed in our simulations that whenever this additional stopping criterion is triggered for a particular shortened node, there typically exists another variable node which upon shortening causes the decoder to converge directly to an error estimate satisfying the original syndrome. Thus, this additional stopping criterion primarily improves latency rather than decoding performance.

As mentioned before, logical errors form a large fraction of decoding failures when shortening to 1, especially at lower error rates. One way to mitigate this is to not set the LLR of the variable node $v$ being shortened to $\lambda_v=-\infty$, but instead to a sufficiently negative finite value. We observed that the choice $\lambda_v = -\ln\left(\frac{1-p}{p}\right)$ provides a good trade-off, since it decreases the percentage of logical errors significantly while maintaining almost the same performance as in Fig. \ref{fig:BB288_0_1}. Decreasing the magnitude of $\lambda_v$ further (while still keeping its sign negative) leads to a further decrease in the number of logical errors, but this comes at the cost of overall decoding performance. This shows that one needs to indeed provide a sufficiently large negative bias, which captures the spirit of code-shortening, in order to get performance improvement.
Another advantage of setting $\lambda_v$ to be sufficiently negative but not $-\infty$ is that the additional stopping criterion mentioned in the paragraph above is not necessary, since sufficiently large messages from neighboring check nodes can make the final message at variable node $v$ positive.

A natural question to ask is whether performance can be improved by shortening variable nodes to both $0$ and $1$. Our observations indicate that this offers little benefit. In particular, it yields only marginal improvement over shortening to 1 alone. This observation again supports the fact that shortening to 1 is particularly favored by decoding dynamics.

Another natural direction to investigate is that of shortening multiple variable nodes. 
In our experiments, we observed that in general, doing so does not improve over the performance of $\mathtt{Minimum\_Weight}/1$ in Fig. \ref{fig:BB288_0_1}. In particular, shortening multiple \emph{randomly selected} variable nodes to 1 on each parallel decoder does not improve performance (on the contrary, the performance degrades since the decoder now has the difficult task of finding an error estimate that yields the received syndrome, and is supported on \emph{all} the shortened variable nodes; note that such an error estimate may not even exist). On the other hand, shortening up to a certain number of variable nodes to 0 improves the performance of $\mathtt{Minimum\_Weight}/0$, but never outperforms $\mathtt{Minimum\_Weight}/1$ in Fig. \ref{fig:BB288_0_1}. However, shortening too many variable nodes to 0 degrades the performance (for similar reasons as before). Thus, shortening randomly chosen variable nodes, either to 1 or to 0, does not outperform the $\mathtt{Minimum\_Weight}/1$ curve in Fig.~\ref{fig:BB288_0_1}. It remains possible that shortening a carefully selected set of variable nodes could yield further performance improvements. However, doing so would require identifying an appropriate set of variable nodes, which would introduce additional search complexity. We leave the investigation of such selection strategies to future work.

\begin{figure}[ht]
    \centering
    \includegraphics[width=\linewidth]{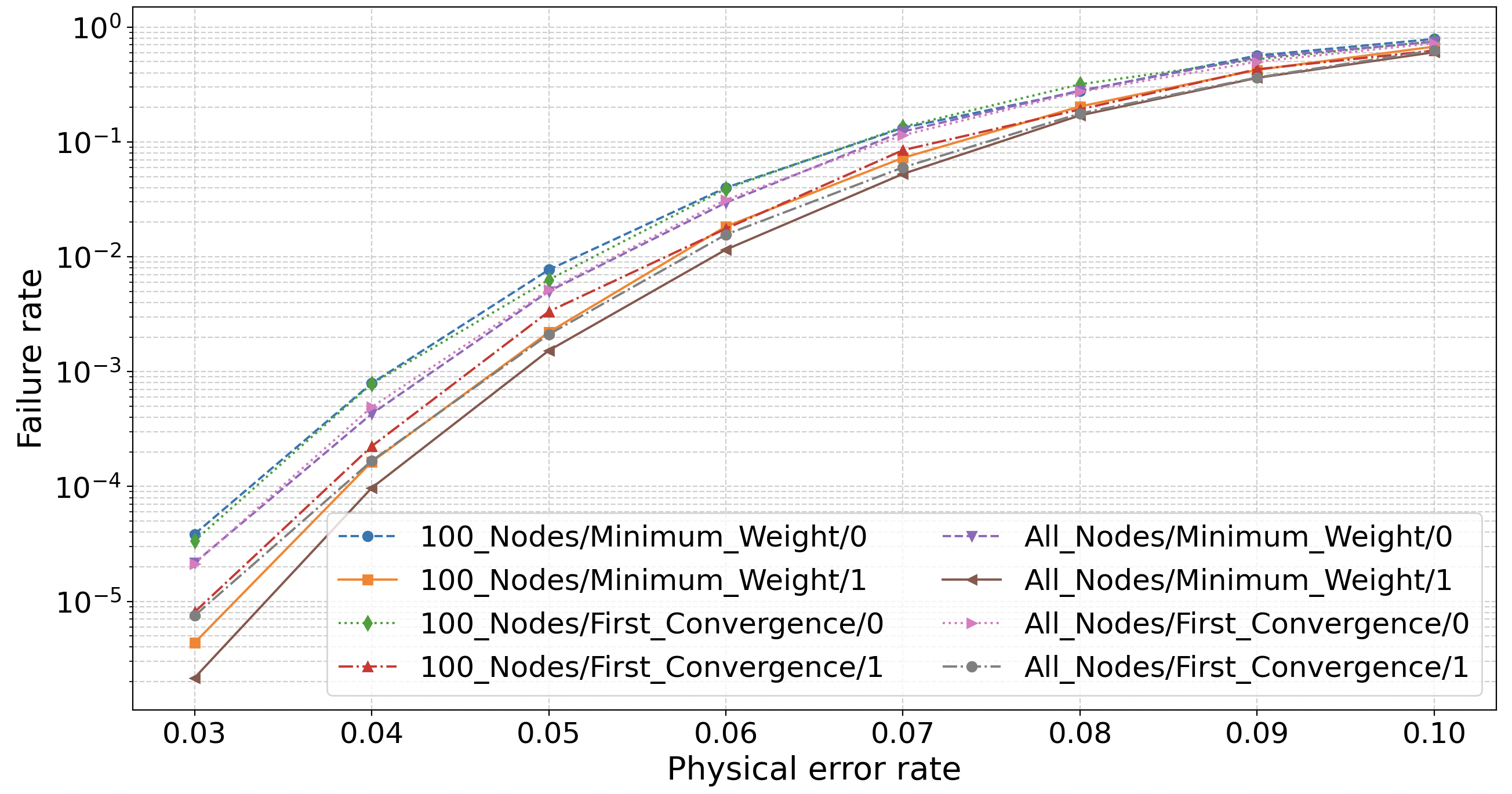}
    \caption{Comparison of impulse decoding (Algorithm~\ref{algo:impulse_decoder}) for the $[[288,12,18]]$ BB code when shortening all $288$ variable nodes versus shortening only the first $100$ variable nodes.}
    \label{fig:BB288_100}
\end{figure}

Following Algorithm \ref{algo:impulse_decoder}, Fig. \ref{fig:BB288_0_1} is obtained by shortening all variable nodes in the Tanner graph. In Fig. \ref{fig:BB288_100}, we compare this with the case where only the first 100 variable nodes (which is about one-third of the total number of variable nodes) are shortened. As is evident from the figure, the difference between shortening all variable nodes versus only the first 100 variable nodes is small. Thus, 100 parallel decoders are sufficient to achieve almost the same performance as in Fig. \ref{fig:BB288_0_1} in practice, especially for the case of shortening to 1.

\begin{figure}[ht]
    \centering
    \includegraphics[width=0.9\linewidth]{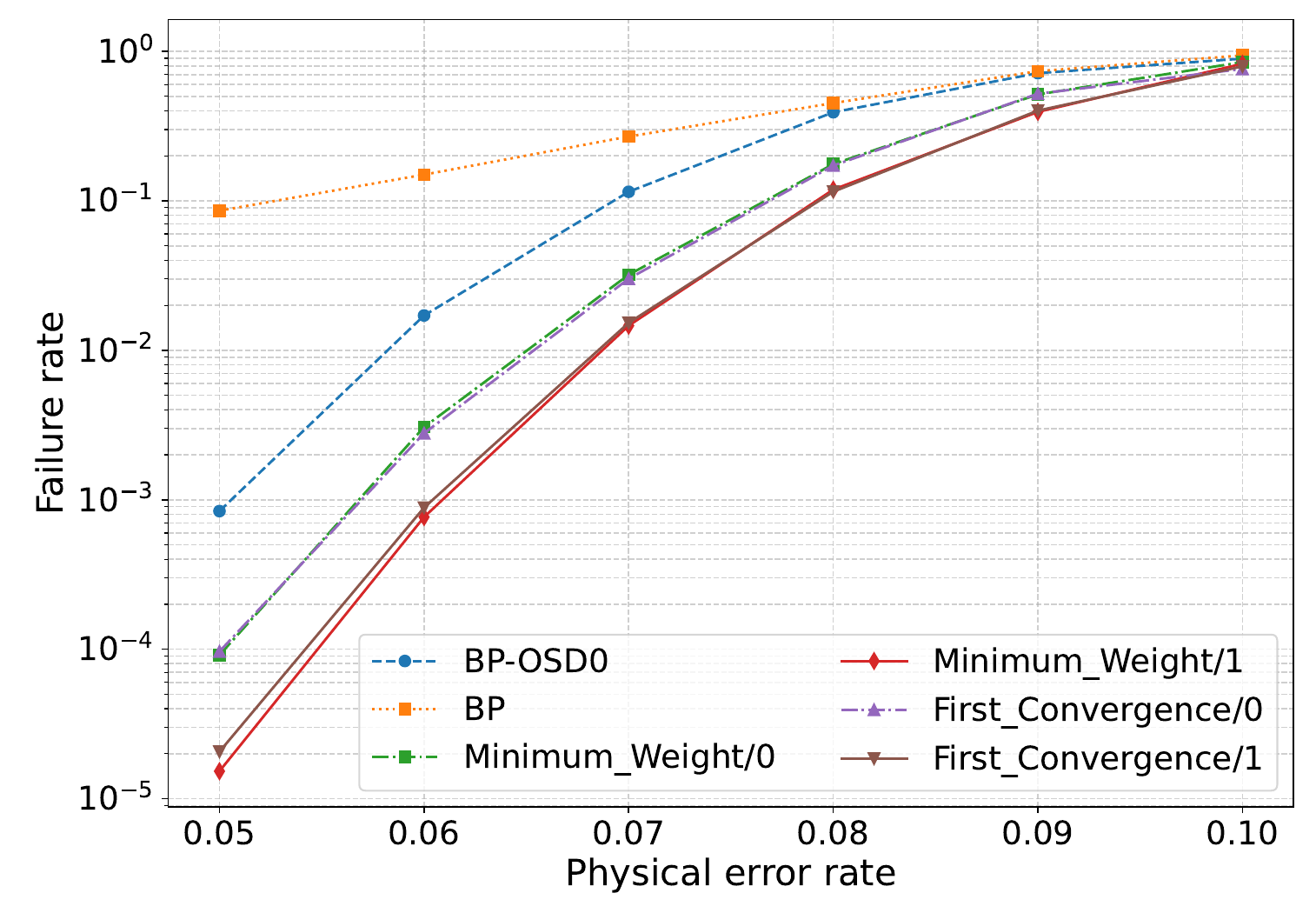}
    \caption{Performance of impulse decoding (Algorithm \ref{algo:impulse_decoder}) for the $[[882,24,18 \le d \le 24]]$ B1 LP code.}
    \label{fig:B1}
\end{figure}

The performance of Algorithm \ref{algo:impulse_decoder} for the $[[882,24,18 \le d \le 24]]$ B1 lifted product (LP) code \cite{panteleev2021degenerate} is shown in Fig. \ref{fig:B1}. It follows the same trends as Fig. \ref{fig:BB288_0_1}.

\begin{figure}[ht]
    \centering
    \includegraphics[width=0.9\linewidth]{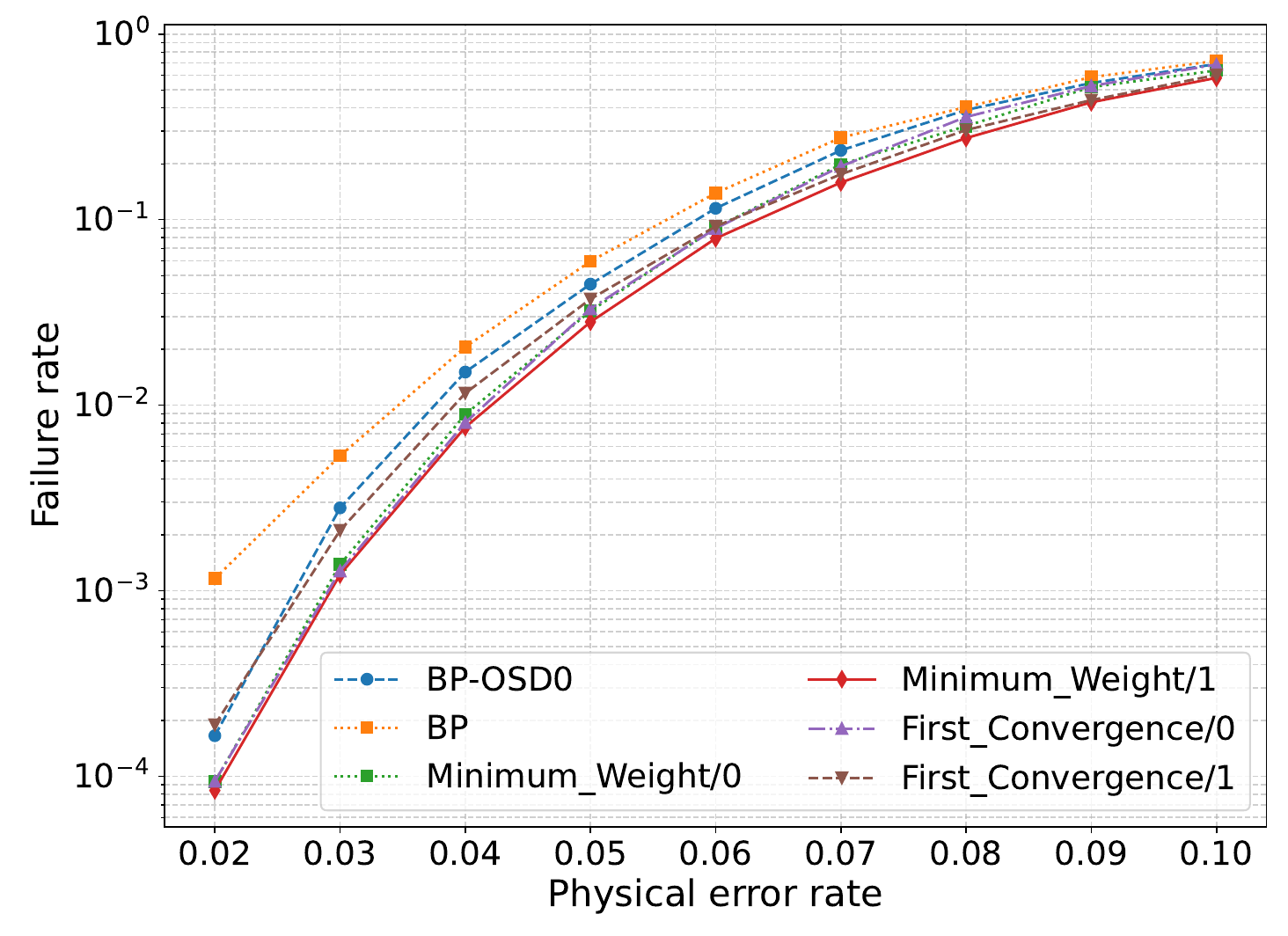}
    \caption{Performance of impulse decoding (Algorithm \ref{algo:impulse_decoder}) for the $[[144,12,12]]$ BB code.}
    \label{fig:BB144_0_1}
\end{figure}

\begin{figure}[htbp]
    \centering

    \begin{subfigure}{0.48\textwidth}
        \centering
        \includegraphics[width=\linewidth]{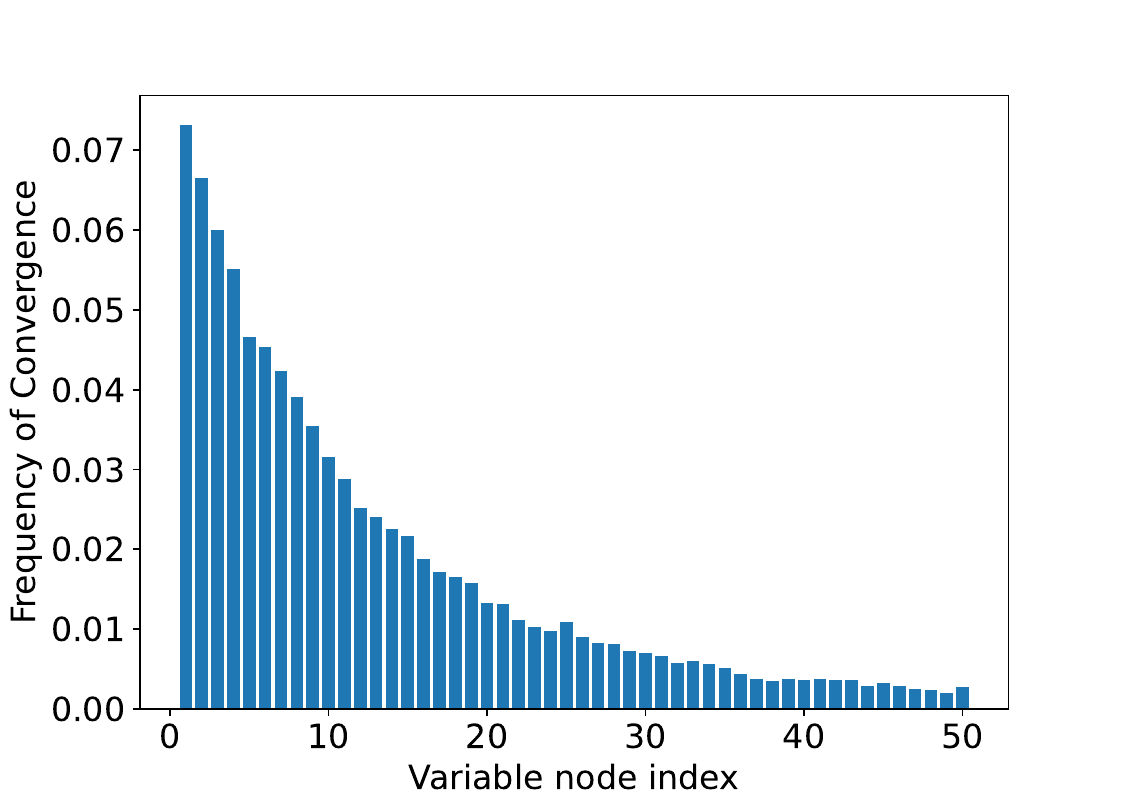}
        \caption{Forward latency.}
        \label{fig:B2_forward_latency}
    \end{subfigure}
    \hfill
    \begin{subfigure}{0.48\textwidth}
        \centering
        \includegraphics[width=\linewidth]{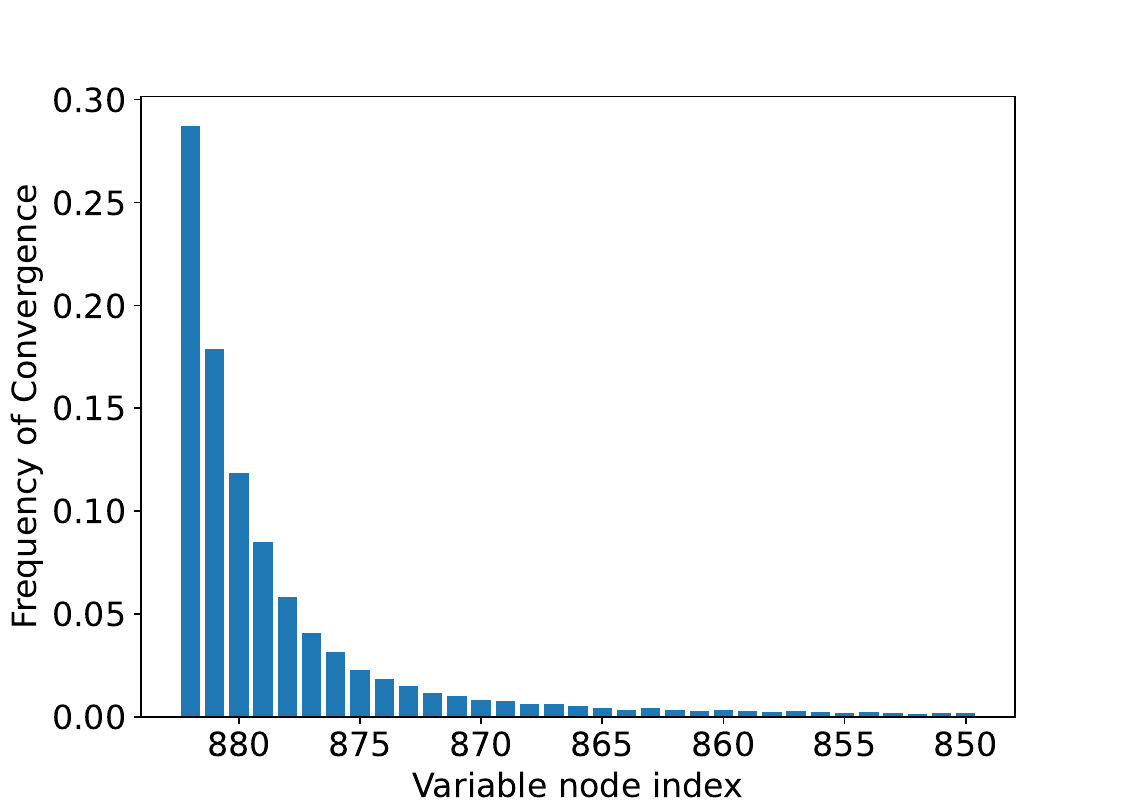}
        \caption{Backward latency.}
        \label{fig:B2_backward_latency}
    \end{subfigure}

    \caption{Frequency of convergence upon shortening variable nodes under $\mathtt{First\_Convergence/1}$ for the $[[882,48,16]]$ B2 LP code  (a) from first to last and (b) from last to first, for a physical error rate $0.03$.}
    \label{fig:B2_latency}
\end{figure}

While the $[[288,12,18]]$ BB code and the $[[882,24,18\le d \le 24]]$ B1 code exhibit similar trends, Fig. \ref{fig:BB144_0_1} shows that the $[[144,12,12]]$ BB code behaves slightly differently. First, the performance improvement over BP-OSD is not as significant. Second, shortening to 0 yields performance comparable to shortening to 1. In fact, under the $\mathtt{First\_Convergence}$ criterion, shortening to 0 performs slightly better than shortening to 1. This is because the $[[144,12,12]]$ BB code exhibits relatively weaker degeneracy. Indeed, for all three codes, the stabilizer generators have weight 6, but the minimum distances of the $[[288,12,18]]$ BB code and the $[[882,24,18\le d\le 24]]$ B1 code are significantly larger (at least three times the stabilizer weight), whereas the distance of the $[[144,12,12]]$ BB code is only twice the stabilizer weight. Consequently, forcing degeneracy in the case of the $[[144,12,12]]$ BB code significantly increases the likelihood of logical errors. By choosing the minimum weight error estimate, logical errors can be avoided, making $\mathtt{Minimum\_Weight}/1$ perform well, as in the case of the other two codes. Moreover, similar to the case of the $[[288,12,18]]$ BB code, the requirement of the number of parallel decoders to achieve virtually the same performance as $\mathtt{Minimum\_Weight}/1$ and $\mathtt{First\_Convergence/1}$ can be reduced from 144 to 50. At the same time, the number of BP iterations in each of the 50 decoders can be reduced from $T=100$ to $T'=30$.

All the codes considered so far were $d_V$-regular, i.e., all variable nodes had the same degree $d_V$. We now discuss the case of an irregular code. The $[[882,48,16]]$ B2 LP code in \cite{panteleev2021degenerate} has the property that the first $441$ variable nodes have degree $3$, while the remaining half of the variable nodes have degree $5$. In Fig. \ref{fig:B2_latency} we compare the latency under $\mathtt{First\_Convergence/1}$ at physical error rate $p=0.03$ for two cases: $(i)$ forward latency: variable nodes are shortened sequentially from first to last (Fig. \ref{fig:B2_forward_latency}), and $(ii)$ backward latency: variable nodes are shortened sequentially from last to first (Fig. \ref{fig:B2_backward_latency}). Recall that exactly half of the stabilizers (of a given type, here we will consider $X$-type stabilizers) are supported on any index $i$. However, this is not true for \emph{low-weight stabilizers}. Assuming that all the low-weight stabilizers appear in $H_X$, the number of low-weight stabilizers supported on index $i$ is precisely the Hamming weight of column $i$, or equivalently, the degree of variable node $i$. Since a BP decoder tries to find the least weight error estimate corresponding to a given syndrome, one can (to first order) assume that upon shortening variable node $v$ of degree $d_v$, the decoder attempts to find one of the $d_v$ degenerate error estimates obtained by adding a low-weight stabilizer supported on variable node $v$ to the channel error. Thus, in the forward latency setting, the decoder first shortens variable nodes of degree $3$ and searches for one of $3$ low-weight error estimates in the shortened code, whereas in the backward latency case, the decoder first shortens variable nodes of degree $5$ and searches for one of $5$ low-weight error estimates in the shortened code. Fig. \ref{fig:B2_latency} shows that the decoder finds it considerably easier to search for an error estimate in the latter case than in the former. Thus, for codes with irregular variable node degree, shortening variable nodes having larger degree before those having smaller degree significantly improves latency, while exhibiting the same performance.

\begin{figure}[ht]
    \centering
    \includegraphics[width=0.9\linewidth]{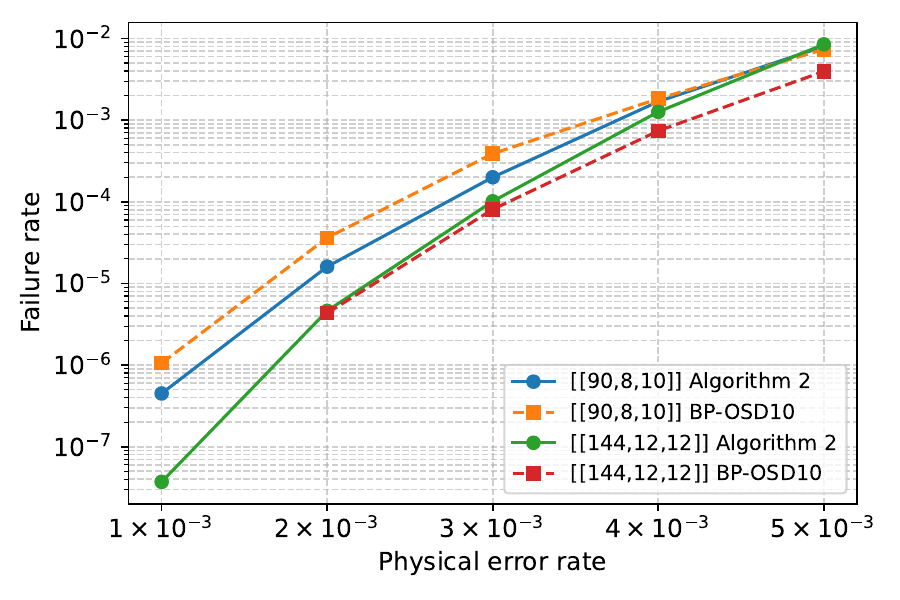}
    \caption{Performance of reliability-based impulse decoding (Algorithm \ref{algo:reliability_impulse_decoder}) under circuit-level noise for the $[[90,8,10]]$ and $[[144,12,12]]$ BB codes.}
    \label{fig:ckt_level}
\end{figure}

\begin{figure}[ht]
    \centering
    \includegraphics[width=0.9\linewidth]{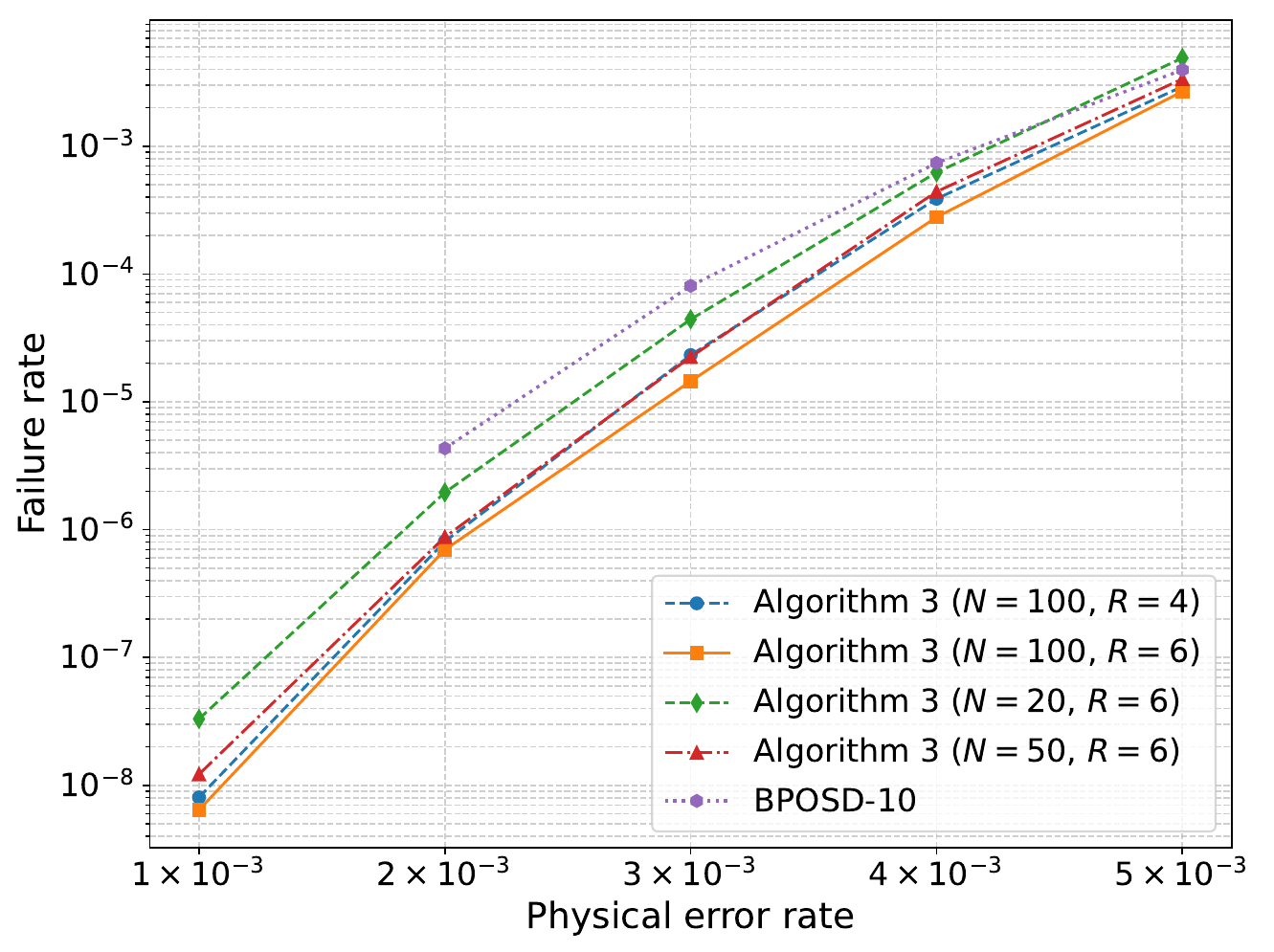}
    \caption{Performance of residual-error-based impulse decoding (Algorithm \ref{algo:residual_error}) for the $[[144,12,12]]$ BB code under circuit-level noise.}
    \label{fig:ckt_level_res_error}
\end{figure}

We now turn to circuit-level noise. We use the min-sum algorithm for this setting, where \eqref{eq:bpchktovar} is replaced by
\begin{align*}
\label{eq:minsum_chktovar}
\begin{split}
\mu_{c \rightarrow v}^{(t)} = \left((-1)^{\un{s}_c}\prod_{v' \in \mathcal{N}(c)\setminus \{v \}} \text{sgn} \left(\nu_{v' \rightarrow c}^{(t-1)}\right) \right )\\
\left ( \underset{v' \in \mathcal{N}(c)\setminus \{v \}}{\text{min}} ~ |\nu_{v' \rightarrow c}^{(t-1)} |\right),
\end{split}
\end{align*}
and $T=100$ message-passing iterations are performed. Failure rate in Fig. \ref{fig:ckt_level} and Fig. \ref{fig:ckt_level_res_error} refers to failure rate per round, and is calculated from the total failure rate $p_\text{tot}$ as $1-(1-p_\text{tot})^{1/{R'}}$, where $R'$ is the number of syndrome measurement rounds \cite{gong2024toward}.

Fig. \ref{fig:ckt_level} shows the performance of Algorithm \ref{algo:reliability_impulse_decoder} for the $[[90,8,10]]$ and $[[144,12,12]]$ BB codes taken from \cite{gong2024toward}. The plots are obtained using $R=2$ and $N=150$ in Algorithm \ref{algo:reliability_impulse_decoder}.
As can be seen from the figure, Algorithm \ref{algo:reliability_impulse_decoder} outperforms BP-OSD10 for the $[[90,8,10]]$ BB code, while achieving performance competitive with BP-OSD10 for the $[[144,12,12]]$ BB code, particularly in the low error-rate regime. We remark here that while Fig. \ref{fig:ckt_level} has been generated using the $\mathtt{Minimum\_Weight}$ criterion, employing the $\mathtt{First\_Convergence}$ criterion leads to almost identical performance.

The performance of Algorithm \ref{algo:residual_error} for various values of $(N,R)$ pairs is depicted in Fig. \ref{fig:ckt_level_res_error}. At least $100$ failures are collected for all physical error rates above $10^{-3}$, while at least $50$ errors are collected for error rate $10^{-3}$.
As Fig. \ref{fig:ckt_level_res_error} shows, using $N=20$ and $R=6$ outperforms BP-OSD10, as well as Algorithm \ref{algo:reliability_impulse_decoder}. As $N$ and $R$ increase, the performance improves further, though with diminishing returns.

\subsection{Complexity Comparison with Prior Approaches}
We end this section with a comparison of the complexity of impulse decoding with some other approaches in the literature. In the $\mathtt{restart\_belief}$ decoder~\cite{valentini2025restartbeliefgeneralquantum}, the first stage involves performing an initial BP run, and, if the decoder does not converge, employing $\eta$ parallel decoders, where decoder $i$ fixes (shortens) the $i$-th least reliable variable node to 1 (analogous to Algorithm~\ref{algo:reliability_impulse_decoder}). However, whenever decoding within a branch does not converge, each parallel decoder flips the least reliable bit (in its respective branch) and restarts the BP procedure, repeating this process up to at most $t=\lfloor \frac{d-1}{2} \rfloor$ times, where $d$ is the minimum distance of the code. For the $[[144,12,12]]$ BB code, the performance of $\mathtt{restart\_belief}$ is illustrated in Fig. 1 in~\cite{valentini2025restartbeliefgeneralquantum}, with the root decoder run for 50 iterations, and branch decoders for 10 iterations, with a total of $\eta = 35$ parallel branches, each re-initialized at most $t=5$ times, for a total of 1800 BP iterations. The $\mathtt{restart\_belief}$ procedure uses one global $\rm argsort$ call (which has complexity $O(n\log(n))$, comparable to BP), and up to $t=5$ $\rm argmin$ calls within each parallel decoder.  
On the other hand, we use Algorithm~\ref{algo:impulse_decoder} for decoding code-capacity noise, which achieves essentially the same performance as $\mathtt{restart\_belief}$ 
(note that physical error rate $p$ in Fig. \ref{fig:BB288_0_1} indicates that $X$-type errors occur with probability $p$, whereas \cite{valentini2025restartbeliefgeneralquantum} employs a depolarizing channel where physical error rate $p$ implies that $X$-type errors occur with probability $2p/3$).
As mentioned before, we can achieve the performance in Fig. \ref{fig:BB144_0_1} by using 50 parallel decoders, each performing 30 BP iterations. Thus, the total number of BP iterations performed in our scheme is $100 + (30 \times 50) = 1600$, which is less than $\mathtt{restart\_belief}$. Further, we do not incur any sorting complexity, except searching for the least weight error estimate, which is also done in the $\mathtt{restart\_belief}$ procedure.
 
The $\mathtt{beam\_search}$ decoder \cite{ye2025beamsearchdecoderquantum} instead performs a tree-based exploration. After the initial BP run, the least reliable bit is fixed once to $0$ and once to $1$, thereby generating two parallel decoding branches. Whenever a branch fails to converge, it is recursively expanded into two additional branches. The performance of $\mathtt{beam\_search}$ for the $[[144,12,12]]$ BB code under circuit-level noise is illustrated in Fig. 2 in~\cite{ye2025beamsearchdecoderquantum}. For such code under circuit-level noise, in comparison, we employ Algorithm~\ref{algo:reliability_impulse_decoder}, with performance illustrated in Fig.~\ref{fig:ckt_level}, which shows a comparable error rate to that of the $\mathtt{beam64\_32res\_640iters}$ configuration. However, the latter requires 20 rounds of 64 parallel BP instances, each run for 30 iterations (except for the first), for a total of 38400 BP iterations; moreover, it requires an $\rm argsort$ call and around 1280 $\rm argmin$ calls, whereas our method consists of 2 rounds of 150 fully parallel decoders, each run for 100 iterations, and only one $\rm argsort$ call, resulting in a lower complexity, especially since all the 150 decoders in each rounds are independent and can run in parallel, while $\mathtt{beam\_search}$ only allows for at most 64 parallel computations.

The highly parallel nature of Algorithm \ref{algo:reliability_impulse_decoder} makes it significantly preferable to the $\mathtt{relayBP}$ decoder \cite{muller2025improved}, which, in contrast, needs to run a fully sequential sequence of $601$ BP instances, each one for a maximum of $60$ iterations, for a total number of $36060$ worst case BP iterations.

We end by noting that Algorithm \ref{algo:residual_error} outperforms Algorithm \ref{algo:reliability_impulse_decoder} with the parameters $N=20, R=6$, which, in the worst case, requires $1+(6\times20)=121$ BP instances, and thus $12100$ BP iterations, along with a single $\rm argsort$ call.

\section{Conclusion}
\label{sec:conclusion}
We demonstrated that degeneracy in CSS codes is closely related to the classical operation of shortening a linear block code, thereby providing a concrete classical interpretation of this phenomenon. Based on this insight, we presented a low-complexity parallel decoding scheme called impulse decoding that achieves state-of-the-art performance under both code-capacity and circuit-level noise.
We then presented another decoding scheme for circuit-level noise based on residual-error decoding. This approach outperformed the first while requiring fewer parallel decoders, at the expense of performing serial processing within each decoder.

We also presented several insights and observations highlighting the interplay between degeneracy and decoder dynamics, and their impact on both performance and latency. In particular, many of these observations indicate that shortening variable nodes to 1 yields significantly better performance and lower latency than shortening to 0.
An important direction for future work, thus, is to identify a suitable set of candidate variable nodes for shortening to 1 (either individually or simultaneously), especially under circuit-level noise, while keeping the additional complexity low. 
Exploring the use of different message-passing schedules, such as serial and layered schedules, is also a practically relevant direction for future work.
Another promising direction is to study the performance of impulse decoding under correlated circuit-level noise \cite{maan2026decoding}, and develop extensions to handle such errors.

\section*{Acknowledgments}
The authors acknowledge the support of the National Science Foundation under grants CIF-2420424, CIF-2106189, CCF-2100013, and CCSS-2052751. We also acknowledge a generous gift from our friends and Maecenases Dora and Barry Bursey. Bane Vasi\'c and Nithin Raveendran have disclosed an outside interest in QEC Labs to the University of Arizona. Conflicts of interest resulting from this interest are being managed by The University of Arizona in accordance with its policies.

\bibliographystyle{ieeetr}
\bibliography{qec_references}

\end{document}